\documentclass[12pt]{article}
\usepackage{multirow}  
\usepackage{amsmath}
\usepackage{amssymb, latexsym}
 \usepackage{amsthm, amsfonts}
 \usepackage{caption}
 \usepackage{natbib}
\usepackage{setspace} 
\usepackage{hyperref}
\usepackage{titlesec}
\usepackage{graphics}
\usepackage{graphicx}
\usepackage{subfig}
\usepackage{rotating}
\titleformat{\section}
    {\normalfont\normalsize\bfseries}
  {\thesection}{8pt}{}
\titleformat{\subsection}
    {\normalfont\normalsize\bfseries}
  {\thesubsection}{5pt}{}
\newtheorem{lemma}{Lemma}
\newtheorem{theorem}{Theorem}
\newcommand\scalemath[2]{\scalebox{#1}{\mbox{\ensuremath{\displaystyle #2}}}}

\def\no{\noindent}
\def\Ytj{Y_{1j}}
\def\Ycj{Y_{0j}}

\def\bx{\boldsymbol{x}} 
\def\bp{\boldsymbol{p}} 
\def\bq{\boldsymbol{q}} 
\def\bs{\mathcal{S}}
\def\bbeta{\boldsymbol{\beta}} 
\def\balpha{\boldsymbol{\alpha}}
\def\blambda{\boldsymbol{\lambda}}
\def\bGamma{\boldsymbol{\Gamma}}
\def\bfeta{\boldsymbol{\eta}}
\def\bA{\boldsymbol{A}}
\def\bB{\boldsymbol{B}}
\def\bv{\boldsymbol{v}}
\def\bV{\boldsymbol{V}}
\def\bW{\boldsymbol{W}}

\def\htau{\hat{\tau}}
\def\hm{\hat{m}}

\def\thetaIPW{\hat{\theta}_{\mbox{\tiny IPW}}}
\def\mutDR{\hat{\mu}_{\mbox{\tiny DR1}}}
\def\mucDR{\hat{\mu}_{\mbox{\tiny DR0}}}
\def\thetaDR{\hat{\theta}_{\mbox{\tiny DR}}}

\def\SELO{\hat{\theta}_{\mbox{\tiny SEL1}}}
\def\mutSELO{\hat{\mu}_{1\mbox{\tiny SEL1}}}
\def\mucSELO{\hat{\mu}_{0\mbox{\tiny SEL1}}}

\def\SELT{\hat{\theta}_{\mbox{\tiny SEL2}}}
\def\mutSELT{\hat{\mu}_{1\mbox{\tiny SEL2}}}

\begin{document}

\centerline{\large {\bf Sample Empirical Likelihood Methods for Causal Inference}}

\bigskip

\centerline{Jingyue Huang, \ Changbao Wu  \ and \ Leilei Zeng \footnote{Jingyue Huang is a doctoral student, Changbao Wu is a Professor, and Leilei Zeng is an Associate Professor, Department of Statistics and  Actuarial Science, University of Waterloo, Waterloo ON N2L 3G1, Canada (E-mails: {\em jingyue.huang@uwaterloo.ca}, \ {\em cbwu@uwaterloo.ca} \ and \ {\em lzeng@uwaterloo.ca}). This research was supported by grants from the Natural Sciences and Engineering Research Council of Canada and the Canadian Statistical Sciences Institute.}}

\bigskip

\bigskip

\hrule

{\small
\begin{quotation}
\no Causal inference is crucial for understanding the true impact of interventions, policies, or actions, enabling informed decision-making and providing insights into the underlying mechanisms that shape our world. 
In this paper, we establish a framework for the estimation and inference of average treatment effects using a two-sample empirical likelihood function. 
Two different approaches to incorporating propensity scores are developed.    
The first approach introduces propensity scores calibrated constraints in addition to the standard model-calibration constraints; 
the second approach uses the propensity scores to form weighted versions of the model-calibration constraints.
The resulting estimators from both approaches are doubly robust. 
The limiting distributions of the two sample empirical likelihood ratio statistics are derived, facilitating the construction of confidence intervals and hypothesis tests for the average treatment effect.
Bootstrap methods for constructing sample empirical likelihood ratio confidence intervals are also discussed for both approaches. 
Finite sample performances of the methods are investigated through simulation studies. 
\vspace{0.3cm}

\no
KEYWORDS \ Causal inference; average treatment effect; sample empirical likelihood methods; model calibration; doubly robust; empirical likelihood ratio statistics.
\end{quotation}
}

\hrule

\bigskip

\section{Introduction}
\label{sec::Intro}
\setcounter{equation}{0}
Causal inference goes beyond observing associations between variables and aims at understanding the underlying causal mechanisms. It answers the question about the causal effect of a treatment, a policy, or an action. Randomized controlled trials have long been considered the gold-standard study design for causal inference in clinical research. However, they are not always feasible due to ethical concerns and budget and time limitations in many situations. Establishing causal inference from observational data provides a feasible alternative but with challenges. The treatment assignments in observational studies are not random and may be influenced by factors associated with outcomes. Such factors are called {\it confounders}, and ignoring them often results in a biased estimator of the causal effect of the treatment. 

Balancing confounders in different treatment groups is essential to causal inference in observational studies. 
This is often achieved by using the propensity score defined as the conditional probability of treatment assignment given a set of covariates \citep{rosenbaum1983ps}.  
Various statistical methods have been proposed based on propensity scores for the estimation of the {\it average treatment effect} (ATE),  including matching \citep{Rosenbaum1985, Abadie2006}, post-stratification \citep{Rosenbaum1984, Rosenbaum1987} and inverse probability weighting (IPW)  \citep{Robins2000, Hirano2003}. 
The consistency of the resulting ATE estimators from the aforementioned methods relies on the correct specification of the propensity score model. 
To protect against the potential model misspecification, the so-called augmented inverse probability weighted (AIPW) estimators \citep{Robins1994, Robins1995} can be constructed by combining inverse probability weighting with outcome regression modelling. 
Such estimators remain consistent if one of the two sets of models, the set of propensity score model and the set of outcome regression models, is misspecified, hence enjoying the property of double robustness. 
In addition, they are more efficient than the standard inverse probability weighted estimators when both sets of models are correctly specified. 

Entropy balancing is an alternative to inverse probability weighting. 
It was first introduced by \cite{Hainmueller2012} and has an attractive finite-sample exact balance property that IPW methods do not have.  
The main idea is to find weights for each observation in the control group that maximize the entropy of these weights, subject to certain calibration constraints. 
The calibration constraints are used to ensure that the weighted covariates' moments in the control group match the corresponding sample moments of the treatment group.  
The entropy balancing method has been shown to be doubly robust with respect to the linear outcome regression and the logistic propensity score regression \citep{Zhao2017}. 
However, the number of constraints required in this approach is usually at least the dimension of the covariates, which can be quite large. 

In this paper, we establish a framework for the estimation and inference of the ATE using a two-sample empirical likelihood function with calibration constraints. Two different approaches are proposed to incorporate the estimated propensity scores.  
One adds constraints on propensity scores along with the standard outcome regression model calibration and parameter constraints;   
the other uses weights based on propensity scores to form a weighted version of the model-calibration and parameter constraints. 
These approaches achieve exact balance in covariates in finite samples with less number of required constraints compared to the entropy balancing, and they generate ATE estimators that are doubly robust.
The asymptotic properties of the sample empirical likelihood (SEL) ratio statistic are established for each approach, enabling the construction of confidence intervals and hypotheses tests for the ATE. 
Bootstrap methods for approximating the distributions of the SEL ratio statistics are also presented.
The proposed methods enjoy the following attractive features shared by general
empirical likelihood-based approaches: (1) problem formulations through a constrained optimization procedure, which
enables incorporations of auxiliary information as additional constraints; and (2) range-respecting and
transformation-invariant properties of the SEL ratio confidence intervals. 

The rest of the paper is organized as follows. Section \ref{sec::setup} provides an introduction to the basic concepts and commonly used methods in causal inference, following the standard potential outcome framework. Sections \ref{sec::SELPSConstraints} and \ref{sec::SELPSWeights} present two different formulations of sample empirical likelihood approaches to causal inference, the former with added constraints on propensity scores and the latter using propensity scores to weight the model-calibration and parameter constraints. 
The point estimator and SEL ratio confidence intervals for the ATE are developed for each approach. Section \ref{sec::simulation} examines the finite sample performance of the proposed methods through simulations. We conclude with a discussion in  Section \ref{sec::discussion}. Proofs and technical details are given in the Appendix.

\section{Causal Inference and Propensity Score Methods}
\label{sec::setup}

The potential outcomes framework is arguably the most widely adopted framework for causal inference.
Under this framework, we define a binary variable $T$ that indicates the treatment assignment so that $T=1$ for treatment and $T=0$ for control, and use  $Y_1$ and $Y_0$ to present the outcomes that would be observed if an individual is under the treatment and control respectively. 
The pair, $(Y_1, Y_0)$, are called potential outcomes since only one of them is observed and the other is a hypothetical latent variable. 
They are used to define causal parameters or estimands.  
A commonly used causal estimand is the average treatment effect (ATE),  defined as the average difference in potential outcomes $\theta=\mu_1-\mu_0$, where $\mu_1=\operatorname{E}(Y_1)$ and $ \mu_0=\operatorname{E}(Y_0)$. 

Consider a random sample $\bs$ drawn from an infinite population. Let $T_j$ and ($\Ytj, \Ycj$) represent the treatment assignment and potential outcomes for a study subject $j$ in this sample, $j=1, \ldots, n$.  
A component $Y_{ij}$ ($i=1$ or $0$) of the pair of potential outcomes is observed if and only if the subject is assigned to the treatment $T_j=i$. 
Thus, the observed outcome is expressed as $Y_j=T_j\Ytj+(1-T_j)\Ycj$, $j\in \bs$. 
Suppose that for each subject $j$, a $p$ dimensional covariate vector $\bx_j$ is also measured. 
The available dataset then comprises the observations $\{(\bx_j, Y_j, T_j); j\in \bs \}$. 
We can split the study subjects by treatment groups such that  $\bs_1=\{j \mid T_j =1; j\in \bs\}$ and $\bs_0=\{j \mid T_j =0; j\in \bs\}$ represent the sub-sample of individuals receiving treatment and control respectively, the available data from each sub-sample are then $\{(\bx_j, \Ytj, T_j=1); j\in \bs_1 \}$ and $\{(\bx_j, \Ycj, T_j=0); j\in \bs_0 \}$ accordingly. 
The sizes of treatment and control groups are $n_1$ and $n_0$, and they should add up to the total sample size $n$. 

In completely randomized studies, the ATE can be simply estimated by the sample mean difference in observed treatment and control outcomes, i.e., $\hat{\theta}=\sum_{j \in \bs} T_j Y_j/n_1-\sum_{j \in \bs}(1-T_j)Y_j/n_0 =\sum_{j \in \bs_1}\Ytj/n_1-\sum_{j \in \bs_0}\Ycj/n_0$.  
This, however, may not be a valid estimator in observational studies. 
The difference in the observed outcomes is not only attributed to the treatment assignments, but also plausibly due to some other covariates that differ between treatment groups, known as the confounders. 
Balancing the confounders in the treatment and control groups is thus essential to causal inference in observational studies. 
A commonly used tool for achieving this is the propensity score (PS), defined as the conditional probability of being assigned treatment given the covariates, 
i.e., $\tau(\bx) = P(T=1\mid \bx)$,
where the subscript $j$ for a subject is omitted.   
Different propensity score methods were developed to remove the effects of confounding when estimating the ATE, including matching, post-stratification and inverse probability weighting. 
The consistency of the resulting estimators relies on two key assumptions outlined by \cite{rosenbaum1983ps}:
\begin{description}
\item[A1.] {\em Strongly Ignorable Treatment Assignment (SITA)}. The treatment assignment and the potential outcomes are independent given the set of covariates, i.e., $(Y_1, Y_0) \perp T \mid \bx$. 
\item[A2.] {\em Positivity Assumption}. The conditional probability of receiving the treatment given the covariates is in the range of $(0,1)$, i.e., $0<P(T=1\mid \bx)<1$ for all $\bx$.
\end{description}
The first condition implies that there are no unmeasured confounders, 
and the second condition ensures that every subject has a strictly positive chance of being in the treatment or the control group. 
Under the first condition, the treatment assignment is ignorable given the propensity score, i.e., $(Y_1, Y_0) \perp T \mid \tau(\bx)$, which enables the construction of consistent estimators for the ATE by conditioning on propensity scores.  

Propensity scores are usually unknown and need to be estimated in observational studies. 
Any method for modelling a binary response variable can be used for this purpose.
For example, using a logistic regression model, the estimated propensity score for a subject $j$ is the predicted probability of treatment derived from the fitted regression model, $\htau_j=\text{expit}(\tilde{\bx}_j^\top\hat{\balpha})=\exp(\tilde{\bx}_j^\top\hat{\balpha})/\{1+\exp(\tilde{\bx}_j^\top\hat{\balpha})\}$, where $\tilde{\bx}_j=(1, \bx_j^\top)^\top$ and $\hat{\balpha}$ is a vector of maximum likelihood estimates of the regression coefficients based on the observed dataset $\{(\bx_j, T_j), j\in \bs\}$. 
One way of using the PS to control for confounding is to weight the observed data. 
The idea is similar to the use of survey sampling weights so that the weighted samples are representatives of specific populations. 
The weight of each individual is calculated as the inverse of the conditional probability of the treatment that the individual actually received to form a synthetic sample in which the distribution of the covariates is independent of the treatment assignment. 
The inverse probability weighted (IPW) estimator of the ATE is then given by  
\[
\thetaIPW= \frac{1}{n} \sum_{j \in \bs_1} \frac{\Ytj}{\htau_j} - \frac{1}{n}  \sum_{j \in \bs_0} \frac{\Ycj}{1-\htau_j}\;. 
\]
By replacing sample size $n$ in the above expression by its consistent estimators $\sum_{j \in \bs_1} \htau_j^{-1}$ and $\sum_{j \in \bs_0} (1 -  \htau_j)^{-1}$ respectively, a H\'{a}jek-type IPW estimator is created that it can be more efficient than the IPW estimator for finite samples \citep{sarndal2003survey}.

The IPW estimators are biased when the propensity score model is misspecified. 
This problem can be mitigated by combining the outcome regression with PS weighting such that the effect estimator is robust to the misspecification of one (but not both) of these models \citep{Robins1994}. 
Consider regression models for two potential outcomes: 
\[
\operatorname{E}(Y_{1}\mid \bx)=m_1(\bx;\bbeta_1) ~~~\mbox{and}~~~\operatorname{E}(Y_{0}\mid \bx)=m_0(\bx;\bbeta_0)\;.
\] 
Under the SITA assumption that $ (Y_1, Y_0) \perp T \mid \bx$, we have $\operatorname{E}(Y_1\mid \bx)=\operatorname{E}(Y_1\mid T=1, \bx)$ and $\operatorname{E}(Y_0\mid \bx)=\operatorname{E}(Y_0\mid T=0, \bx)$. This implies that $\bbeta_1$ and $\bbeta_0$ can be estimated by fitting the above outcome regression models to the observed data of the treatment group and control group, $\{(\bx_j,\Ytj), j\in\bs_1\}$ and $\{(\bx_j, \Ycj), j\in\bs_0\}$, separately. 
The resulting parameter estimates are used to calculate the predicted values $\hm_{1j}=m_1(\bx_j; \hat{\bbeta}_1)$ and $\hm_{0j}=m_0(\bx_j; \hat{\bbeta}_0)$ for the potential outcomes $Y_{1j}$ and $Y_{0j}$ for each individual $j$ in the whole sample. 
Having estimated the PS, $Y_{1}$ and $Y_{0}$, we combine these values to construct the doubly robust (DR) estimators of the mean response in the presence and absence of treatment as 
\begin{eqnarray*}
\mutDR = \frac{1}{n} \sum_{j \in \bs} \left [Y_{1j} + \frac{(Y_{1j} -\hm_{1j})(T_j -\htau_{j}) }{\htau_{j}} \right]= \frac{1}{n} \sum_{j \in \bs_1} \frac{ Y_{1j}- \hm_{1j}}{\htau_{j}} + \frac{1}{n} \sum_{j \in \bs}^{n}\hm_{1j} \;, \\
\mucDR = \frac{1}{n} \sum_{j \in \bs} \left [Y_{0j} - \frac{(Y_{0j} -\hm_{0j})(T_j -\htau_{j}) }{1-\htau_{j}} \right]
=\frac{1}{n} \sum_{j \in \bs_0} \frac{ Y_{0j}- \hm_{0j}}{1-\htau_{j}} + \frac{1}{n} \sum_{j \in \bs}^{n}\hm_{0j}  \,. 
\end{eqnarray*}
The above estimators are represented as the quantity of interest (the response if everyone had been exposed/unexposed) plus a second term referred to as the ``augmentation". This component is formed as the product of two bias terms, one from the PS model and the other from the outcome regression model. If either bias term equals zero, then it ``zeros out" the other nonzero bias term from the incorrect model. This explains the doubly robust property. 
The difference between the estimated means, 
$\thetaDR= \mutDR - \mucDR$, 
is the DR estimate of the ATE, which is also referred to as the augmented IPW estimator. 

\section{SEL with Constraints on Propensity Scores}
\label{sec::SELPSConstraints}

\subsection{Empirical Likelihood}
\label{subsec::PEL and SEL}
Empirical likelihood (EL) was first proposed by \cite{Owen1988} and is a general non-parametric method for making inference. 
It uses the likelihood principle in a way that is analogous to the usual parametric likelihood without making any assumptions about the underlying distribution. 
The general framework of EL methods involves maximizing the empirical likelihood function subject to the normalization constraint and the constraint for the parameter of interest.
The EL ratio statistic is defined as the ratio of the ``restricted" maximum empirical likelihood under a specific parameter value to the ``global" maximum empirical likelihood under any parameter value. 
For independent and identically distributed random samples, 
\cite{Owen1988} showed that the EL ratio for the population mean has a standard limiting chi-squared distribution and used this result to obtain confidence intervals. 
Generalizations to multivariate functionals and linear regression models were proposed in \cite{Owen1990} and \cite{Owen1991}, respectively. 
\cite{Qin1994} linked the empirical likelihood with estimating equations, and derived asymptotic properties of the maximum EL estimator and the EL ratio statistic when the constraints for the parameter of interest are based on a set of unbiased estimating equations. 
In the context of complex surveys, the survey weights are incorporated into the construction of the empirical likelihood for the estimation of a finite population mean \citep{Chen1999, Wu2006} through the so-called pseudo empirical likelihood (PEL). 
Alternatively, the standard empirical likelihood function is used while the sampling weights are incorporated into the constraints, leading to the so-called sample empirical likelihood (SEL) \citep{Chen2014, Zhao2022}.
General frameworks and methodologies of both approaches in the context of analysis of complex survey data are comprehensively reviewed in \cite{Zhao2019}.

For causal inference problems, \cite{huang2024pseudoempirical} developed PEL-based methods for the estimation and inference of the ATE, where the survey weights in the PEL function were replaced with weights based on propensity scores. 
This paper concerns the use of sample empirical likelihood instead and proposes two SEL-based approaches that are different in how propensity scores are incorporated into the constraints.  
The first approach follows the standard empirical likelihood formulation for independent sample data, but uses additional constraints based on the estimated propensity scores. It is the focus of the remainder of this section. 
The other SEL approach uses propensity scores to weight the model-calibration and parameter constraints. It is described in detail in Section \ref{sec::SELPSWeights}. 

\subsection{Maximum SEL Estimator of the ATE}
\label{subsec::SELOPOINT}

The idea of incorporating the extra propensity score calibration constraints besides model-calibration constraints into the standard framework of EL was described by \cite{Han2013} for multiply robust estimation in the context of missing data problems.
Here, we adopt this idea to obtain a doubly robust estimator for the ATE for causal inference problems with two samples. 
Let $\bp_i = (p_{i1},\cdots,p_{in_i})^\top$ be a set of discrete probability measure imposed over $\bs_i$ for $i=0,1$. We consider a joint two-sample SEL function in the standard log-form 
\begin{equation}
\ell_{\mbox{\tiny SEL}}(\bp_1, \bp_0) = \sum_{j\in \bs_1}\log(p_{1j})+\sum_{j\in \bs_0}\log(p_{0j})
\label{eq::jointSEL}
\tag{$l_p$}
\end{equation}
subject to the following constraints: 
\begin{equation}
    \sum_{j\in\bs_1}p_{1j}=1  \;\;\;{\rm and} \;\;\; \sum_{j\in\bs_0}p_{0j}=1 \;, \hskip 2.2cm 
    \tag{$C1$}
    \label{eq::jnorm}    
\end{equation}
\begin{equation}
    \sum_{j \in \bs_1} p_{1j}\hm_{1j}=\bar{\hm}_1 \;\;\;{\rm and} \;\;\;  \sum_{j \in \bs_0} p_{0j}\hm_{0j}=\bar{\hm}_0 \;,\hskip .5cm
    \label{eq::jcal}
    \tag{$C2$}
\end{equation}
\begin{equation}
        \sum_{j\in\bs_1}p_{1j}\htau_j=\Bar{\htau} \;\;\;{\rm and }\;\;\; 
        \sum_{j\in\bs_0}p_{0j}(1-\htau_j)=1-\Bar{\htau} \;,
    \tag{$C3$}
    \label{eq::jPS}
\end{equation}
where $\bar{\hm}_1 = n^{-1}\sum_{j\in\bs}\hm_{1j}$, $\bar{\hm}_0 = n^{-1}\sum_{j\in\bs}\hm_{0j}$, and $\Bar{\htau} = n^{-1}\sum_{j \in \bs}\htau_j$. 
As one can see, in addition to the normalization constraints (\ref{eq::jnorm}) and model-calibration constraints (\ref{eq::jcal}) that are typically involved in EL methods, we introduce constraints (\ref{eq::jPS})  on propensity scores such that the weighted averages of the propensity scores in $\bs_1$ and in $\bs_0$ are forced to agree to their corresponding overall sample means. This is in line with the idea of covariates balance through propensity scores. Note that the second part of (\ref{eq::jPS}) is equivalent to $\sum_{j\in\bs_0}p_{0j} \htau_j = \Bar{\htau}$ under (\ref{eq::jnorm}). 
Let $(\hat{\bp}_1^\top, \hat{\bp}_0^\top) = (\hat{p}_{11},\cdots,\hat{p}_{1n_1}, \hat{p}_{01},\cdots,\hat{p}_{0n_0})$ be the maximizer of the joint SEL function (\ref{eq::jointSEL}) under constraints (\ref{eq::jnorm}), (\ref{eq::jcal}) and  (\ref{eq::jPS}). 
The maximum SEL estimators for the mean of the potential outcomes are  
$\hat{\mu}_{i\mbox{\tiny SEL1}} =\sum_{j\in \bs_i}\hat{p}_{ij} Y_{ij}\;,\; i=0, 1$, 
and the estimator of the ATE is given by $\SELO=\mutSELO-\mucSELO$.

It turns out that finding $(\hat{\bp}_1^\top, \hat{\bp}_0^\top)$ amounts to solving two one-sample optimization problems using standard techniques. 
For each sample $\bs_i$, $i=0, 1$, we have a one-sample SEL function 
$\ell_{\mbox{\tiny SEL}}(\bp_i) = \sum_{j\in \bs_i}\log(p_{ij})$
associated with a set of constraints 
\[
\sum_{j \in \bs_i} p_{ij}=1 \;\; \mbox{and}\; \; \sum_{j \in \bs_i} p_{ij} \hat{\bv}_{ij} = \boldsymbol{0}\, ,
\]
where $\hat{\bv}_{ij}=(\htau_j-\Bar{\htau}, \hm_{ij}-\bar{\hm}_i)^\top$. 
By uisng the Lagrange multiplier method, we obtain $\hat{p}_{ij}=1/\{n_i(1+\hat{\blambda}_i^\top\hat{\bv}_{ij})\}$, where $\hat{\blambda}_i$ is the solution to 
$\sum_{j\in \bs_i} \hat{\bv}_{ij} / (1+\blambda_i^\top\hat{\bv}_{ij})=\boldsymbol{0}$.
However, the $\hat{\blambda}_i$ does not have the usual order of $O_p(n^{-1/2})$ even when the propensity score model is correctly specified, which makes it difficult to show the consistency of the proposed maximum SEL estimator following the standard route for empirical likelihood. 

One way to circumvent the aforementioned technical issue is to do a suitable reformulation of the constrained maximization problem. Let $q_{1j}=p_{1j}\htau_j/\Bar{\htau}$ for $j\in\bs_1$ and $q_{0j}=p_{0j}(1-\htau_j)/(1-\Bar{\htau})$ for $j\in\bs_0$. Maximizing $\ell_{\mbox{\tiny SEL}}(\bp_1, \bp_0)$ of (\ref{eq::jointSEL}) subject to constraints  (\ref{eq::jnorm}), (\ref{eq::jcal}) and (\ref{eq::jPS})  is equivalent to maximizing the likelihood function
\begin{equation}
\ell_{\mbox{\tiny SEL}}(\bq_1, \bq_0) = \sum_{j\in \bs_1}\log(q_{1j})+\sum_{j\in \bs_0}\log(q_{0j})
\label{eq::jointSELq}
\tag{$l_q$}
\end{equation}
under the constraints 
\begin{equation}
   \begin{aligned}
        &\sum_{j\in\bs_1}q_{1j}=1\,, \;\;\;\;\;\;\;\;\;\;\;\;\;\;\;\;\;\;\;\;\;\;\;\;\;\;\;\;\;
        \sum_{j\in\bs_0}q_{0j}=1\,,\\
        &\sum_{j\in\bs_1}q_{1j}\htau_j^{-1}\left(\htau_j-\bar{\htau}\right)=0\,,\;\;\;\;\;\;\;\;\;\;\;
        \sum_{j\in\bs_0}q_{0j}\left(1-\htau_j\right)^{-1}\left\{\left(1-\htau_j\right)-\left(1-\bar{\htau}\right)\right\}=0\,,\\
        &\sum_{j \in \bs_1} q_{1j}\htau_j^{-1}(\hm_{1j}-\bar{\hm}_1)=0\,,\;\;\;\;\;\;\;
        \sum_{j \in \bs_0} q_{0j}\left(1-\htau_j\right)^{-1}(\hm_{0j}-\bar{\hm}_0)=0 \; .
    \end{aligned}
    \tag{$C_q1$}
    \label{eq::SELjointconsq}
\end{equation}
The values of $\bq_0$ and $\bq_1$ that maximize the SEL function (\ref{eq::jointSELq}) under the constraints (\ref{eq::SELjointconsq}) are given by 
$\hat{q}_{ij}=1/\{n_i(1+\htau_j^{-1}\hat{\blambda}_{iq}^\top\hat{\bv}_{ij})\}$ for $j\in\bs_i$, $i=1, 0$, where the Lagrange multipliers $\hat{\blambda}_{1q}$ and $\hat{\blambda}_{0q}$ are the solutions to
\begin{equation}
  \sum_{j \in \bs_1}\frac{\htau_j^{-1} \hat{\bv}_{1j}}{1+\htau_j^{-1}\blambda_{1q}^\top\hat{\bv}_{1j}}=\boldsymbol{0}\;\;\;\; \mbox{and}\;\;\;\;  
  \sum_{j \in \bs_0}\frac{(1-\htau_j)^{-1} \hat{\bv}_{0j}}{1+(1-\htau_j)^{-1}\blambda_{0q}^\top\hat{\bv}_{0j}}=\boldsymbol{0} \;.
  \label{eq::SELLagequation}
\end{equation}
The maximum SEL estimators for $\mu_i$ remain the same, i.e., $\hat{\mu}_{1\mbox{\tiny SEL1}} = \sum_{j\in \bs_1}\hat{p}_{1j} Y_{1j}=
    \sum_{j\in \bs_1}{\Bar{\htau}}\hat{q}_{1j}Y_{1j}/{\htau_j}$ and $\hat{\mu}_{0\mbox{\tiny SEL1}} = \sum_{j\in \bs_0}\hat{p}_{0j} Y_{0j}=
    \sum_{j\in \bs_0}{(1-\Bar{\htau})}\hat{q}_{0j}Y_{0j}/(1-{\htau_j})$. 
It can be shown that the order of $\hat{\blambda}_{iq}$, $i=1, 0$, is now $O_p(n^{-1/2})$  under the assumed propensity score model, which facilitates the development of asymptotic properties for $\hat{\mu}_{1\mbox{\tiny SEL1}}$ and $\hat{\mu}_{0\mbox{\tiny SEL1}}$. 
The results on $\mutSELO$ are stated in Theorem \ref{thm::SEL1mu1} below, with the detailed proof given in Section \ref{sec::ProofDRmut}.  Similar results hold for  $\mucSELO$ as well as the maximum SEL estimator $\SELO=\mutSELO-\mucSELO$ for the ATE. 

\begin{theorem}
\label{thm::SEL1mu1}
Under certain regularity conditions, the maximum SEL estimator $\mutSELO$ is doubly robust with respect to the propensity score and outcome regression models.
Moreover, if the propensity score model is correctly specified, we have $\hat{\mu}_{1\mbox{\tiny SEL1}}=\mu_1^0+O_p(n^{-1/2})$ where $\mu_1^0$ represents the true value of $\mu_1$. 
\end{theorem}

\subsection{SEL Ratio Confidence Intervals}
\label{sec::SELR1}

A major advantage of the proposed SEL approach is that it is fairly straightforward to make inference about the parameter of interest using the SEL ratio statistic. 
By deriving the limiting distribution of the SEL ratio statistic, we can construct confidence intervals and perform hypothesis tests for the parameter of interest, i.e., the ATE $\theta$, through the inclusion of an additional parameter constraint 
\begin{equation}
    \sum_{j\in \bs_1}p_{1j}Y_{1j}-\sum_{j\in \bs_0}p_{0j}Y_{0j}=\theta \;. 
    \tag{$C4$}
    \label{eq::para}
\end{equation}
Let $\hat{\bp}^\top_0(\theta)=(\hat{p}_{01}(\theta)$, $\cdots, \hat{p}_{0n_0}(\theta))$ and $\hat{\bp}^\top_1(\theta)=(\hat{p}_{11}(\theta), \cdots, \hat{p}_{1n_1}(\theta))$ denote the ``restricted" maximizer of $\ell_{\mbox{\tiny SEL}}(\bp_1, \bp_0)$ 
under the constraints (\ref{eq::jnorm})-(\ref{eq::jPS}) plus the additional parameter constraint (\ref{eq::para}) for a given $\theta$. 
We refer to $\hat{\bp}_0$ and $\hat{\bp}_1$ that maximize $\ell_{\mbox{\tiny SEL}}(\bp_1, \bp_0)$ under the constraints (\ref{eq::jnorm})-(\ref{eq::jPS}) the ``global" maximizer. 
The profile sample empirical likelihood function for the parameter $\theta$ is given by  
\begin{equation}
\ell_{\mbox{\tiny SEL}}(\hat{\bp}_1(\theta), \hat{\bp}_0(\theta)) = \sum_{j\in \bs_1}\log(\hat{p}_{1j}(\theta))+\sum_{j\in \bs_0}\log(\hat{p}_{0j}(\theta))\;. 
    \label{eq::SELR1pl}
\end{equation}
Recognize that the maximum SEL estimator $\SELO$ derived in the previous section maximizes the profile SEL function $\ell_{\mbox{\tiny SEL}}(\hat{\bp}_1(\theta), \hat{\bp}_0(\theta))$, and $\hat{\bp}_i = \hat{\bp}_i(\SELO)$ for $i=0, 1$. 
The SEL ratio function for $\theta$ is therefore given by 
$
r_{\mbox{\tiny SEL}}(\theta)=\ell_{\mbox{\tiny SEL}}(\hat{\bp}_1(\theta), \hat{\bp}_0(\theta))-\ell_{\mbox{\tiny SEL}}(\hat{\bp}_1, \hat{\bp}_0)
$.

Once again, a reformulation of the constraints is required for the derivation of the limiting distribution of the SEL ratio statistic for the same reason as mentioned in Section \ref{subsec::SELOPOINT}. 
The parameter constraint (\ref{eq::para}) can be re-written as 
\begin{equation}
    \sum_{j \in \bs_1} q_{1j}\htau_j^{-1}(Y_{1j}-\mu_1)=0\;\;\;{\rm and}\;\;\;\sum_{j \in \bs_0} q_{0j}\left(1-\htau_j\right)^{-1}(Y_{0j}-\mu_1+\theta)=0 \; ,
\tag{$C_q2$}
\label{eq::paraq}
\end{equation}
where $q_{1j}$ and $q_{0j}$ are defined as before, and $\mu_1$ is introduced as a nuisance parameter for computational purpose. 
The optimization problem now becomes maximizing the SEL function $\ell_{\mbox{\tiny SEL}}(\bq_1, \bq_0)$ under the constraints listed in (\ref{eq::SELjointconsq}) and (\ref{eq::paraq}). 
Note that the inclusion of a nuisance parameter $\mu_1$ in the parameter constraint makes it impossible to segregate the problem into two stand-alone one-sample problems, hence adding complexity to the current problem. 
Let $\hat{\bq}_i(\mu_1, \theta)$, $i=1, 0$, denote the ``restricted" maximizer of $\ell_{\mbox{\tiny SEL}}(\bq_1, \bq_0)$ for a given $(\mu_1, \theta)$. 
The profile SEL function for $(\mu_1, \theta)$ is $\ell_{\mbox{\tiny SEL}}(\hat{\bq}_1(\mu_1, \theta), \hat{\bq}_0(\mu_1, \theta))$,  maximizing it over $\mu_1$ for a given $\theta$ leads to $\hat{\mu}_1(\theta)$. 
We present two lemmas on the computational equivalence with the nuisance parameter $\mu_1$. 

\begin{lemma}
The maximum SEL estimators $\mutSELO$ and $\SELO$ maximize the profile SEL function 
$\ell_{\mbox{\tiny SEL}}(\hat{\bq}_1(\mu_1, \theta), \hat{\bq}_0(\mu_1, \theta))$ with respect to $(\mu_1, \theta)$. Moreover, the maximum SEL estimator $ \SELO$ maximizes the profile SEL function $\ell_{\mbox{\tiny SEL}}(\hat{\bq}_1(\hat{\mu}_1(\theta), \theta), \hat{\bq}_0(\hat{\mu}_1(\theta), \theta))$.    
\end{lemma}

\begin{lemma}
The sample empirical likelihood ratio function $r_{\mbox{\tiny SEL1}}(\theta) = \ell_{\mbox{\tiny SEL}}\left(\hat{\bq}_1\left(\theta\right), \hat{\bq}_0\left(\theta\right)\right)-\ell_{\mbox{\tiny SEL}}\left(\hat{\bq}_1, \hat{\bq}_0\right)$ for $\theta$ can be equivalently computed as 
\[
r_{\mbox{\tiny SEL1}}(\theta)
    =\ell_{\mbox{\tiny SEL}}\left(\hat{\bq}_1\left(\hat{\mu}_1\left(\theta\right), \theta\right), \hat{\bq}_0\left( \hat{\mu}_1\left(\theta\right), \theta\right)\right)-\ell_{\mbox{\tiny SEL}}\left(\hat{\bq}_1\left(\mutSELO, \SELO\right), \hat{\bq}_0\left(\mutSELO, \SELO\right)\right).
\]
\end{lemma}

Existing algorithms for empirical likelihood can be used to compute the joint profile SEL function $\ell_{\mbox{\tiny SEL}}(\hat{\bq}_1(\mu_1, \theta), \hat{\bq}_0(\mu_1, \theta))$ with given values of $(\mu_1, \theta)$. 
Note that the constraints in (\ref{eq::SELjointconsq}) and (\ref{eq::paraq}) can be re-written as 
\begin{equation}
    \begin{aligned}
        &\sum_{j\in\bs_1}q_{1j}=1\,, \;\;\;\;\;\;\;\;\;\;\;\;\;\;\;\;\;\;\;\;\;\;\;
        \sum_{j\in\bs_0}q_{0j}=1\,,\\
        &\sum_{j\in\bs_1}q_{1j}\boldsymbol{g}_{1j}(\mu_1)=\boldsymbol{0}\,,\;\;\;\;\;\;\;\;\;\;\;
        \sum_{j\in\bs_0}q_{0j}\boldsymbol{g}_{0j}(\mu_1, \theta)=\boldsymbol{0}\,,
    \end{aligned}
    \label{eq::SELjointconsqv2}
    \tag{$C_q3$}
\end{equation}
where $\boldsymbol{g}_{1j}(\mu_1)=\htau_j^{-1}(\htau_j-\bar{\htau}, \hm_{1j}-\bar{\hm}_1, \Ytj-\mu_1)^\top$ and 
$\boldsymbol{g}_{0j}(\mu_1, \theta)=(1-\htau_j)^{-1}\{(1-\htau_j)-(1-\bar{\htau}), \hm_{0j}-\bar{\hm}_0, Y_{0j}-\mu_1+\theta\}^\top$.
For fixed $(\mu_1, \theta)$, maximizing $\ell_{\mbox{\tiny SEL}}(\bq_1, \bq_0)$ with respect to $\bp_1$ and $\bp_0$ under the constraints in (\ref{eq::SELjointconsqv2}) leads to the solution 
\[
\hat{q}_{1j}(\mu_1)=\frac{1}{n_1\left(1+\hat{\blambda}_1^\top\boldsymbol{g}_{1j}(\mu_1)\right)}\;\;\;{\rm and }\;\;\;\hat{q}_{0j}(\mu_1,\theta)=\frac{1}{n_0\left(1+\hat{\blambda}_0^\top\boldsymbol{g}_{0j}(\mu_1, \theta)\right)}\, ,
\]
where the Lagrange multipliers $\hat{\blambda}_1$ and $\hat{\blambda}_0$ are obtained by solving 
\begin{equation}
 \sum_{j \in \bs_1}\frac{\boldsymbol{g}_{1j}(\mu_1)}{1+\blambda_1^\top\boldsymbol{g}_{1j}(\mu_1)}=\boldsymbol{0} \;\;\;{\rm and }\;\;\;\sum_{j \in \bs_0}\frac{\boldsymbol{g}_{0j}(\mu_1, \theta)}{1+\blambda_0^\top\boldsymbol{g}_{0j}(\mu_1, \theta)}=\boldsymbol{0}\, .
  \label{eq::SELJointLagEqua} 
\end{equation}
Since $\mutSELO$ and $\SELO$ maximize the profile SEL function
$\ell_{\mbox{\tiny SEL}}(\hat{\bq}_1(\mu_1, \theta), \hat{\bq}_0(\mu_1, \theta))$, their limiting distribution can be derived based on the Taylor expansion of the profile likelihood score. 
We present the following main results and leave the details in the proofs for the theorems given later. 
For clarification of presentation, we let $\mu_1^0$ and $\theta^0$ denote the true values of $\mu_1$ and $\theta$, $\boldsymbol{W}_{11}$ and $\boldsymbol{W}_{12}$ be the limit values of  $n^{-1}\sum_{j \in \bs_1}\boldsymbol{g}_{1j}(\mu_1^0)\boldsymbol{g}_{1j}^\top(\mu_1^0)$ and $n^{-1}\sum_{j \in \bs_0}\boldsymbol{g}_{0j}(\mu_1^0,\theta^0)\boldsymbol{g}_{0j}^\top(\mu_1^0, \theta^0)$ respectively, and the matrix $\boldsymbol{\Omega}_1$ represent the asymptotic variance-covariance matrix of the vector  
\[
\frac{1}{\sqrt{n}}\sum_{j \in \bs}\begin{pmatrix}
T_j\boldsymbol{g}_{1j}(\mu_1^0)\\
(1-T_j)\boldsymbol{g}_{0j}(\mu_1^0, \theta^0)
\end{pmatrix}\;. 
\]
Define $\bA$ as a $2\times2$ block matrix taking the following form
\[
\bA=\begin{pmatrix}
\bA_{11} & \bA_{12}\\
\bA_{12}^\top & \boldsymbol{0}
\end{pmatrix}\, ,
\]
where 
\[
\bA_{11}=\begin{pmatrix}
\boldsymbol{W}_{11} & \boldsymbol{0}\\
\boldsymbol{0} & \boldsymbol{W}_{12} 
\end{pmatrix}\, ,
\;\;\; 
\bA_{12}=\begin{pmatrix}
\bGamma_1 & \boldsymbol{0}\\
\bGamma_1 & -\bGamma_1
\end{pmatrix}\;,
\;\;\; \mbox{and} \; \; \; 
\bGamma_1=(0, 0, 1)^\top\;.
\]
Its inverse can be calculated using the inverse formula for $2\times 2$ block matrices. Let $\bA^{12}$ and $\bA^{11}$ denote the $[1, 2]$ and $[1, 1]$ block sections of $\bA^{-1}$ respectively. 
The limiting distribution of the maximum SEL estimators $\mutSELO$ and $\SELO$ is given in Theorem \ref{thm::SEL1normalPoint} below. 

\begin{theorem}
\label{thm::SEL1normalPoint}
Let $\hat{\bfeta}_{\mbox{\tiny SEL1}} = (\mutSELO, \SELO)^\top$ and $\bfeta^{0}=(\mu_1^{0}, \theta^{0})^\top$. Under certain regularity conditions and a correctly specified propensity score model, we have $\sqrt{n}\left(\hat{\bfeta}_{\mbox{\tiny SEL1}}-\bfeta^0\right)\stackrel{d}{\rightarrow}\operatorname{MVN}(\boldsymbol{0}, \bV_1)$ as $n\rightarrow\infty$, where $\bV_1=[\bA^{12} ]^\top\boldsymbol{\Omega}_1\bA^{12}$.
\end{theorem} 

The asymptotic properties of the SEL ratio statistic $r_{\mbox{\tiny SEL1}}(\theta)$ are presented in Theorem \ref{thm::SEL1chi2}. 
Additional quantities are required to aid the exposition of this theorem.
In particular, we define a matrix 
\[
\bB=\begin{pmatrix}
\bB_{11} & \bB_{12}\\
\bB_{12}^\top & 0
\end{pmatrix}\,
\]
where $\bB_{11}=\bA_{11}$ and $\bB_{12}=(\bGamma_1^\top, \bGamma_1^\top)^\top$. 
Again we calculate its inverse and use $\bB^{11}$ to denote the $[1, 1]$ block section of $\bB^{-1}$. 
Note that the rank of the matrix $\boldsymbol{\Omega}_1^{1/2}\left(\bB^{11}-\bA^{11}\right)\boldsymbol{\Omega}_1^{1/2}$ is one.
Furthermore, let $\chi^2_1$ denote a chi-squared distribution with one degree of freedom.

\begin{theorem}
\label{thm::SEL1chi2}
Under certain regularity conditions and a correctly specified propensity score model, the scaled sample empirical likelihood ratio statistic $-2r_{\mbox{\tiny SEL1}}(\theta)/\delta_1$ with $\theta=\theta^0$ converges in distribution to a $\chi^2_1$-distributed random variable,
where $\delta_1$ is the non-zero eigenvalue of the matrix $\boldsymbol{\Omega}_1^{1/2}\left(\bB^{11}-\bA^{11}\right)\boldsymbol{\Omega}_1^{1/2}$. 
\end{theorem}

The proofs for both Theorem \ref{thm::SEL1normalPoint} and Theorem \ref{thm::SEL1chi2} are given in Section \ref{sec::ProofSEL1Point} and Section \ref{sec::ProofSEL1chi2} respectively. 
In practice, $\bW_{11}$ and $\bW_{12}$ can be replaced by their corresponding plug-in estimators.
To facilitate the derivation of the above two theorems and the estimation of the variance-covariance matrix $\boldsymbol{\Omega}_1$, we introduce a vector $\boldsymbol{b}_j$, which is given by
\begin{equation}\scalemath{0.9}{
\boldsymbol{b}_j=\left(\begin{array}{c}
(T_j-\tau_j^0)-(T_j / \tau_j^0-1)\operatorname{E}(\tau_j^0)+\left(T_j-\tau_{j}^{0}\right) (\boldsymbol{G}+\boldsymbol{H})\boldsymbol{D}^{-1} \tilde{\bx}_j \\ 
(T_j/\tau_{j}^{0}-1)\left(m_{1j}^{*}-\operatorname{E}(m_{1j}^*)\right)+\left(T_j-\tau_{j}^{0}\right) \boldsymbol{J}\boldsymbol{D}^{-1} \tilde{\bx}_j \\ 
T_j\left(Y_{1 j}-\mu_{1}^{0}\right)/\tau_{j}^{0}-\left(T_j-\tau_{j}^{0}\right) \boldsymbol{KD}^{-1} \tilde{\bx}_j \\
\{1-({1-T_j})/({1-\tau_{j}^{0}})\}(1-\operatorname{E}(\tau_j^0))+({\tau_j^0-T_j})-\left(T_j-\tau_{j}^0\right) (\boldsymbol{L}-\boldsymbol{H})\boldsymbol{D}^{-1} \tilde{\bx}_j \\ 
\left\{({1-T_j})/(1-\tau_j^0)-1\right\}\left(m_{0 j}^{*}-\operatorname{E}({m}_{0j}^{*})\right)-\left(T_j-\tau_{j}^{0}\right) \boldsymbol{MD}^{-1} \tilde{\bx}_j \\ 
(1-T_j)\left(Y_{0 j}-\mu_{1}^{0}+\theta^{0}\right)/({1-\tau_{j}^{0}})-\left(T_j-\tau_{j}^{0}\right) \boldsymbol{ND}^{-1} \tilde{\bx}_j\end{array}\right)\,. 
\label{eq::bj}
}
\end{equation}
In the expression of $\boldsymbol{b}_j$, ${\tau}_j^0=\tau(\bx_j;\balpha^0)$ denotes the propensity score for subject $j$ under the true PS model with $\balpha^0$ representing the true value of $\balpha$;  $m_{ij}^{*} = m_i(\bx_j, \bbeta_i^{*})$ for $i=0, 1$, where $\bbeta_i^{*}$ is the probability limit of $\hat{\bbeta}_i$ under the assumed outcome regression model; 
and 
\begin{equation*}
    \begin{aligned}
        &\boldsymbol{D}=-\operatorname{E}\left[\tau_{j}^{0}\left(1-\tau_{j}^{0}\right) \tilde{\bx}_j \tilde{\bx}_j^{\top}\right], \;\;\;\; \; \; \; \; \; \; \; \; \; \; \; \; \; \; 
        \boldsymbol{G}=-\operatorname{E}(\tau_j^0)\operatorname{E}\left[\left(1-\tau_{j}^{0}\right) \tilde{\bx}_j^{\top}\right],\\
        &\boldsymbol{H}=\operatorname{E}\left[\tau_{j}^{0}\left(1-\tau_{j}^{0}\right)\tilde{\bx}_j^{\top}\right], \;\;\;\;\;\;\;\;\;\;\; \; \; \; \; \; \; \; \; \; \; \; \; \; \; 
        \boldsymbol{J}=\operatorname{E}\left[\left(m_{1 j}^{*}-\operatorname{E}(m_{1j}^*)\right)(1-\tau_j^0)\tilde{\bx}_j^{\top}\right],\\
        &\boldsymbol{K}=-\operatorname{E}\left[T_j\left(1-\tau_{j}^{0}\right) \left(\Ytj-\mu_1^0\right)\tilde{\bx}_j^{\top}/\tau_j^0\right],
        \boldsymbol{L}=(1-\operatorname{E}(\tau_j^0))\operatorname{E}\left[\tau_{j}^{0} \tilde{\bx}_j^{\top}\right], \;\;\;\;\;\\
        &\boldsymbol{M}=\operatorname{E}\left[\tau_{j}^{0} \left(m_{0 j}^{*}-\operatorname{E}({m}_{0j}^{*}\right)\tilde{\bx}_j^\top\right], \; \; \; \; \; \; \; \; \; \; \; \; 
        \boldsymbol{N} =\operatorname{E}\left[(1-T_j)\tau_{j}^{0} \left(Y_{0j}-\mu_1^0+\theta^0\right)\tilde{\bx}_j^\top/\left(1-\tau_j^0\right)\right].
    \end{aligned}
\end{equation*}
The rationales and details of obtaining $\boldsymbol{b}_j$ can be found in Section \ref{sec::ProofSEL1Point}. 
Finally, the variance-covariance matrix $\boldsymbol{\Omega}_1$ can be estimated by $\hat{\boldsymbol{\Omega}}_1=n^{-1}\sum_{j \in \bs}(\hat{\boldsymbol{b}}_j-\bar{\hat{\boldsymbol{b}}})(\hat{\boldsymbol{b}}_j-\bar{\hat{\boldsymbol{b}}})^\top$, where $\bar{\hat{\boldsymbol{b}}}=n^{-1}\sum_{j\in\bs}\hat{\boldsymbol{b}}_j$, and $\hat{\boldsymbol{b}}_j$ represents the plug-in estimator of $\boldsymbol{b}_j$.


Using the result from Theorem \ref{thm::SEL1chi2}, a $100(1-\alpha)\%$ confidence interval for $\theta$ can be constructed as
$\{\theta \mid -2r_{\mbox{\tiny SEL1}}(\theta)/\hat{\delta}_1\leq \chi^2_1(\alpha)\}$, 
where $\hat{\delta}_1$ is the non-zero eigenvalue of $\hat{\boldsymbol{\Omega}}_1^{1/2}(\hat{\boldsymbol{B}}^{11}-\hat{\bA}^{11})\hat{\boldsymbol{\Omega}}_1^{1/2}$, with $\hat{\bA}^{11}$ and $\hat{\boldsymbol{B}}^{11}$ calculated based on the plug-in estimators for the matrics $\bA$ and $\boldsymbol{B}$.
In terms of computation, the constrained maximization problem is reformulated as a stratified sampling format which enables the direct application of algorithms and R codes from \cite{Wu2005} and \cite{Wu2006}.

\subsection{Bootstrap Approach for SEL Ratio Confidence Intervals}
\label{sec::bootstrapSEL}

Bootstrap provides an alternative approach for approximating the distribution of the SEL ratio statistic.  
It allows us to determine the limiting distribution of the SEL ratio statistic through the empirical distribution derived from the bootstrap samples, circumventing the complex procedures associated with computing the scaling constant $\delta_1$ involved in the limiting distribution. 
We outline the steps of a proposed bootstrap procedure for constructing the SEL ratio confidence intervals for the ATE as follows. 
\begin{description}
    \item[Step 1] Select the $b$-th bootstrap sample $\bs^{[b]}$ of $n$ units from the initial sample $\bs$ using simple random sampling with replacement. Let $\bs^{[b]}_1=\{j \mid j\in\bs^{[b]}\text{ and }T_j=1\}$ and $\bs^{[b]}_0=\{j \mid j\in\bs^{[b]}\text{ and }T_j=0\}$. 
    
    \item[Step 2]Fit the propensity score model to the $b$-th bootstrap sample $\bs^{[b]}$ to obtain estimates $\hat{\balpha}^{[b]}$ and propensity scores $\htau_j^{[b]}=\tau(\bx_j; \hat{\balpha}^{[b]})$ for $j \in \bs^{[b]}$. Fit the outcome regression model to each of the two bootstrap subsamples $\bs^{[b]}_1$ and $\bs^{[b]}_0$, obtain estimates $\hat{\bbeta}_1^{[b]}$ and $\hat{\bbeta}_0^{[b]}$ respectively, and calculate the estimated potential outcomes $\hm_{1j}^{[b]}=m_1(\bx_j; \hat{\bbeta}_1^{[b]})$ and $\hm_{0j}^{[b]}=m_0(\bx_j; \hat{\bbeta}_0^{[b]})$ for all the units $j\in \bs^{[b]}$.
    
    \item[Step 3] Construct the bootstrap version of the SEL function as
    \begin{equation*}
      \ell^{[b]}_{\mbox{\tiny SEL}}(\bp_1, \bp_0) = \sum_{j\in \bs^{[b]}_1}\log(p_{1j})+\sum_{j\in \bs_0^{[b]}}\log(p_{0j})
    \end{equation*}
    and specify the bootstrap versions of the constraints in (\ref{eq::jnorm})-(\ref{eq::para}) as 
    \begin{equation*}
        \begin{aligned}
        &\sum_{j\in\bs_1^{[b]}}p_{1j}=1\,, \;\;\;\;\;\;\;\;\;\;\;\;\;\;\;\;\;\;\;\;\;\;\;\;\;\;\;
        \sum_{j\in\bs_0^{[b]}}p_{0j}=1\,,\\
        &\sum_{j\in\bs_1^{[b]}}p_{1j}\htau_j^{[b]}=\frac{1}{n}\sum_{j\in\bs^{[b]}}\htau_j^{[b]}\,,\;\;\;\;\;\;\;\;\;
        \sum_{j\in\bs_0^{[b]}}p_{0j}\left(1-\htau_j^{[b]}\right)=\frac{1}{n}\sum_{j\in\bs^{[b]}}\left(1-\htau_j^{[b]}\right)\,,\\
        &\sum_{j \in \bs_{1}^{[b]}} p_{1j}\hm_{1j}^{[b]}=\frac{1}{n}\sum_{j\in\bs^{[b]}}\hm_{1j}^{[b]}\,,\;\;\;\;\;\;\;\sum_{j \in \bs_{0}^{[b]}} p_{0j}\hm_{0j}^{[b]}=n^{-1}\sum_{j\in\bs^{[b]}}\hm_{0j}^{[b]}\,,\\
        &\sum_{j \in \bs_{1}^{[b]}} p_{1j}Y_{1j}-\sum_{j \in \bs_{0}^{[b]}} p_{0j}Y_{0j}=\theta\, .
        \end{aligned}
    \end{equation*}
     Compute the SEL ratio $r^{[b]}_{\mbox{\tiny SEL1}}(\theta)$ using $\ell^{[b]}_{\mbox{\tiny SEL}}(\bp_1, \bp_0)$ and above constraints at $\theta=\SELO$. 
    
   \item[Step 4]  Repeat Steps 1-3 a total of $B$ times to obtain $r^{[b]}_{\mbox{\tiny SEL1}}(\theta)$ at $\theta=\SELO$ for $b=1, \dots, B$.
   
\end{description}

The number $B$ is commonly chosen as $1000$. 
Let $b_\alpha$ denote the lower $100\alpha$-th quantile of the sequence $r_{\mbox{\tiny SEL1}}^{[b]}(\SELO)$, $b=1,\dots,B$, obtained in Step 4. 
A bootstrap-calibrated $(1-\alpha)$-level SEL ratio confidence interval for $\theta$ can be calculated by
\begin{equation}
    \mathcal{C}=\left\{\theta \mid r_{\mbox{\tiny SEL1}}(\theta)>b_{\alpha}\right\}\, ,
    \notag
\end{equation}
where $r_{\mbox{\tiny SEL1}}(\theta)$ represents the SEL ratio function based on the original sample.

\section{SEL with Propensity Score Weighted Constraints}
\label{sec::SELPSWeights}
\subsection{SEL Estimation of the ATE}
\label{sec::SELpoint2}

Another way to incorporate the propensity scores into the SEL-based approach is to use weighted versions of the model-calibration and parameter constraints based on the estimated propensity score.  
The idea is similar to the SEL approach to design-based survey data analysis where a standard EL function was defined for the sample and the design weights were treated as an intrinsic part of the estimating functions for the parameter of interest \citep{Chen2014, Zhao2022}.  
In our context with two samples, $\bp_1$ and $\bp_0$ are the vectors of the discrete probability measure assigned to samples $\bs_0$ and $\bs_1$. 
We consider maximizing the joint SEL function
\begin{equation*}
\ell_{\mbox{\tiny SEL}}(\bp_1, \bp_0) = \sum_{j\in \bs_1}\log(p_{1j})+\sum_{j\in \bs_0}\log(p_{0j})\;, 
\end{equation*}
subject to the weighted versions of constraints 
\begin{equation}
    \begin{aligned}
    &\sum_{j\in\bs_1}p_{1j}=1  \;, \hskip 4cm   \sum_{j\in\bs_0}p_{0j}=1 \;,  \\
    & \sum_{j \in \bs_{1}} p_{1j}\htau_j^{-1}\left(\hm_{1j}-\bar{\hm}_1\right)=0 \;, \hskip 1.2cm  \sum_{j \in \bs_{0}} p_{0j}(1-\htau_j)^{-1}\left(\hm_{0j}-\bar{\hm}_0\right)=0\;. 
    \end{aligned}
    \tag{$C_{\mbox{\tiny SEL2}}$}
    \label{eq::CSEL2}
\end{equation}
The solutions to this constrained maximization problem are $\hat{p}_{1j}=1/ \{n_1[1+\hat{\lambda}_1\htau_1^{-1}(\hm_{1j}-\bar{\hm}_1)]\}$ for $j\in \bs_1$ and $\hat{p}_{0j}=1/\{n_0[1+\hat{\lambda}_0(1-\htau_j)^{-1}(\hm_{0j}-\bar{\hm}_0)]\}$ for $j\in \bs_0$, where the scalar Lagrange multipliers $\hat{\lambda}_1$ and $\hat{\lambda}_0$ solve the equations
\[
\sum_{j \in \bs_1}\frac{\hat{\tau}_j^{-1}(\hm_{1j}-\bar{\hm}_1)}{1+\lambda_1\htau_j^{-1}(\hm_{1j}-\bar{\hm}_1)}=0 ~~~\mbox{and}~~~
\sum_{j \in \bs_0}\frac{(1-\hat{\tau}_j)^{-1}(\hm_{0j}-\bar{\hm}_0)}{1+\lambda_0(1-\htau_j)^{-1}(\hm_{0j}-\bar{\hm}_0)}=0\;. 
\] 
The maximum SEL estimators for $\mu_1$ and $\mu_0$, denoted by $\hat{\mu}_{1\mbox{\tiny SEL2}}$ and $\hat{\mu}_{0\mbox{\tiny SEL2}}$, are the solutions to the weighted estimating equations 
$$
\sum_{j \in \bs_{1}} \hat{p}_{1j}\htau_j^{-1}\left(Y_{1j}-\mu_1\right)=0 ~~~\mbox{and}~~~ \sum_{j \in \bs_{1}} \hat{p}_{0j}(1-\htau_j)^{-1}\left(Y_{0j}-\mu_0\right)=0\;. 
$$
The above equations can also be viewed as PS weighted parameter constraints to be included in (\ref{eq::CSEL2}). 
Finally, we obtain the maximum SEL estimator of the ATE given by $\hat{\theta}_{\mbox{\tiny SEL2}}= \hat{\mu}_{1\mbox{\tiny SEL2}} -\hat{\mu}_{0\mbox{\tiny SEL2}}$. 
Following the same arguments used for Theorem \ref{thm::SEL1mu1} in Section \ref{subsec::SELOPOINT}, we can argue that $\hat{\mu}_{1\mbox{\tiny SEL2}}$ is doubly robust with its asymptotic properties presented formally in the theorem below. 
The same results hold for $\hat{\mu}_{0\mbox{\tiny SEL2}}$ and $\hat{\theta}_{\mbox{\tiny SEL2}}$. 
\begin{theorem}
Under certain regularity conditions,  the estimator $\hat{\mu}_{1\mbox{\tiny SEL2}}$ is doubly robust in the sense that it is consistent if either the propensity score or the outcome regression model is correctly specified.
Additionally, we have $\hat{\mu}_{1\mbox{\tiny SEL2}}=\mu_1^0+O_p(n^{-1/2})$ when the propensity score model is correctly specified.
\end{theorem}

\subsection{SEL Ratio Confidence Intervals}

We integrate the parameter of interest, the ATE $\theta$, into the proposed SEL procedure for the purpose of establishing the asymptotic properties of the SEL ratio statistic for $\theta$.  
The optimization problem in section \ref{sec::SELpoint2} is equivalent to the maximization of $\ell_{\mbox{\tiny SEL}}(\bp_1, \bp_0)$ under constraints 
\begin{equation}
    \begin{aligned}
        &\sum_{j\in\bs_1}p_{1j}=1\,, \;\;\;\;\;\;\;\;\;\;\;\;\;\;\;\;\;\;\;\;\;\;\;\;\;\;\;\;\;
        \sum_{j\in\bs_0}p_{0j}=1\,,\\
        &\sum_{j \in \bs_{1}} p_{1j}\htau_j^{-1}(\hm_{1j}-\bar{\hm}_1)=0\,,\;\;\;\;\;\;\;\sum_{j \in \bs_{0}} p_{0j}\left(1-\htau_j\right)^{-1}(\hm_{0j}-\bar{\hm}_0)=0\,,\\
        &\sum_{j \in \bs_{1}} p_{1j}\htau_j^{-1}(Y_{1j}-\mu_1)=0\,,\;\;\;\;\;\;\;\;\;\sum_{j \in \bs_{0}} p_{0j}\left(1-\htau_j\right)^{-1}(Y_{0j}-\mu_1+\theta)=0\,, 
    \end{aligned}
    \tag{$C_{\mbox{\tiny SEL2}}$}
    \label{eq::SELjointcons2}  
\end{equation}
where the extra parameter constraints listed at the bottom are essentially estimating equations for parameters $\mu_1$ and $\theta$. 
For any given $(\mu_1, \theta)$, maximizing $\ell_{\mbox{\tiny SEL}}(\bp_1, \bp_0)$ with respect to $\bp_1$ and $\bp_1$ under the constraints gives the  ``restricted" maximum SEL estimators $\hat{\bp}_0(\mu_1, \theta)$ and $\hat{\bp}_1(\mu_1, \theta)$. 
The profile SEL function for parameters $(\mu_1, \theta)$ is then $\ell_{\mbox{\tiny SEL}}(\hat{\bp}_1(\mu_1, \theta), \hat{\bp}_0(\mu_1, \theta))$, and the value of $\mu_1$ that maximizes this function at a given $\theta$ is denoted by $\hat{\mu}_1(\theta)$. 
A bit of logical deduction leads to the following lemmas. 

\begin{lemma}
The maximum SEL estimators $\hat{\mu}_{1\mbox{\tiny SEL2}}$ and $\hat{\theta}_{\mbox{\tiny SEL2}}$ obtained in Section \ref{sec::SELpoint2} also maximize the profile likelihood $\ell_{\mbox{\tiny SEL}}(\hat{\bp}_1(\mu_1, \theta), \hat{\bp}_0(\mu_1, \theta))$. 
In addition, $\hat{\bp}_1 = \hat{\bp}_1(\hat{\mu}_{1\mbox{\tiny SEL2}}, \hat{\theta}_{\mbox{\tiny SEL2}})$ and $\hat{\bp}_0=\hat{\bp}_0(\hat{\mu}_{1\mbox{\tiny SEL2}}, \hat{\theta}_{\mbox{\tiny SEL2}})$, where $\hat{\bp}_1$ and $\hat{\bp}_0$ are the ``global" maximum SEL estimators obtained in Section \ref{sec::SELpoint2}. 
\end{lemma}
\begin{lemma}
The SEL ratio function for the ATE $\theta$ is given by 
$
r_{\mbox{\tiny SEL2}}\left(\theta\right)
= \ell_{\mbox{\tiny SEL}}(\hat{\bp}_1\left(\hat{\mu}_1\left(\theta\right),\theta\right)$, $\hat{\bp}_0\left(\hat{\mu}_1\left(\theta\right),\theta\right))-\ell_{\mbox{\tiny SEL}}\left(\hat{\bp}_1, \hat{\bp}_0\right)
$,
which is equivalently given by 
$
r_{\mbox{\tiny SEL2}}\left(\theta\right)
=\ell_{\mbox{\tiny SEL}}(\hat{\bp}_1\left(\hat{\mu}_1\left(\theta\right),\theta\right)$, $\hat{\bp}_0\left(\hat{\mu}_1\left(\theta\right),\theta\right))-\ell_{\mbox{\tiny SEL}}\left(\hat{\bp}_1\left(\mutSELT, \SELT\right), \hat{\bp}_0\left(\mutSELT, \SELT\right)\right)    
$.
\end{lemma}

The asymptotic properties of the SEL ratio statistic are evaluated by using the same techniques employed for the first approach in Section \ref{sec::SELR1}. Let $\boldsymbol{g}_{1j}^S(\mu_1)=\htau_j^{-1}(\hm_{1j}-\bar{\hm}_1, \Ytj-\mu_1)^\top$ and 
$\boldsymbol{g}_{0j}^S(\mu_1, \theta)=(1-\htau_j)^{-1}( \hm_{0j}-\bar{\hm}_0, Y_{0j}-\mu_1+\theta)^\top$, the constraints in (\ref{eq::SELjointcons2}) is then expressed in a more compact form
\begin{equation}
    \begin{aligned}
        &\sum_{j\in\bs_1}p_{1j}=1\,, \;\;\;\;\;\;\;\;\;\;\;\;\;\;\;\;\;\;\;\;\;\;\;
        \sum_{j\in\bs_0}p_{0j}=1\,,\\
        &\sum_{j\in\bs_1}p_{1j}\boldsymbol{g}_{1j}^S(\mu_1)=\boldsymbol{0}\,,\;\;\;\;\;\;\;\;\;\;\;
        \sum_{j\in\bs_0}p_{0j}\boldsymbol{g}_{0j}^S(\mu_1, \theta)=\boldsymbol{0}\,. 
    \end{aligned}
    \notag
\end{equation}
The standard Lagrange multiplier methods are applied to calculate the ``restricted" maximum SEL estimators for $\bp_1$ and $\bp_1$ at a given $(\mu_1, \theta)$. We have 
\[
\hat{p}_{1j}(\mu_1)=\frac{1}{n_1\left(1+\hat{\blambda}_1^\top\boldsymbol{g}_{1j}^S(\mu_1)\right)} 
~~~\mbox{and}~~~\hat{p}_{0j}(\mu_1,\theta)=\frac{1}{n_0\left(1+\hat{\blambda}_0^\top\boldsymbol{g}_{0j}^S(\mu_1, \theta)\right)}\, ,
\]
where $\hat{\blambda}_1$ and $\hat{\blambda}_0$ are the solutions to
\begin{equation}
 \sum_{j \in \bs_1}\frac{\boldsymbol{g}_{1j}^S(\mu_1)}{1+\blambda_1^\top\boldsymbol{g}_{1j}^S(\mu_1)}=\boldsymbol{0} \;\;\;{\rm and }\;\;\;\sum_{j \in \bs_0}\frac{\boldsymbol{g}_{0j}^S(\mu_1, \theta)}{1+\blambda_0^\top\boldsymbol{g}_{0j}^S(\mu_1, \theta)}=\boldsymbol{0}\, .
  \notag
\end{equation}
The limits of  $n^{-1}\sum_{j \in \bs_1}\boldsymbol{g}_{1j}^S(\mu_1^0)\{\boldsymbol{g}_{1j}^S(\mu_1^0)\}^\top$ and $n^{-1}\sum_{j \in \bs_0}\boldsymbol{g}_{0j}^S(\mu_1^0, \theta^0)\{\boldsymbol{g}_{0j}^S(\mu_1^0, \theta^0)\}^\top$ are denoted by $\boldsymbol{W}_{21}$ and $\boldsymbol{W}_{22}$ respectively. 
We define a $2\times 2$ block matrix $\boldsymbol{E}$ in the form 
\[
\boldsymbol{E}=\begin{pmatrix}
\boldsymbol{E}_{11} & \boldsymbol{E}_{12}\\
\boldsymbol{E}_{21}^\top & \boldsymbol{0}
\end{pmatrix}\, ,
\]
where 
\[
\boldsymbol{E}_{11}=\begin{pmatrix}
\boldsymbol{W}_{21} & \boldsymbol{0}\\
\boldsymbol{0} & \boldsymbol{W}_{22} 
\end{pmatrix}\, ,
\;\;\; 
\boldsymbol{E}_{12}=\begin{pmatrix}
\bGamma_2 & \boldsymbol{0}\\
\bGamma_2 & -\bGamma_2
\end{pmatrix}\, ,
\;\;\; {\rm and} \;\;\;
\bGamma_2=(0,1)^\top\,,
\]
and a block matrix $\boldsymbol{F}$ such that 
\[
\boldsymbol{F}=\begin{pmatrix}
\boldsymbol{E}_{11} & \boldsymbol{F}_{12}\\
\boldsymbol{F}_{12}^\top & \boldsymbol{0}
\end{pmatrix}\,. 
\]
The inverse formula for $2\times2$ block matrices is used to obtain $\boldsymbol{E}^{-1}$ and $\boldsymbol{F}^{-1}$.
We represent the $[i, j]$ block section of these matrices by $\boldsymbol{E}^{ij}$ and $\boldsymbol{F}^{ij}$ accordingly. 
Let $\boldsymbol{\Omega}_2$ denote the variance-covariance matrix of the vector
\[
\boldsymbol{b}_{2j}=\left(\begin{array}{c} (T_j/\tau_{j}^{0}-1)\left\{m_{1j}^{*}-\operatorname{E}({m}_{1j}^{*})\right\}+\left(T_j-\tau_{j}^{0}\right) \boldsymbol{JD}^{-1} \tilde{\bx}_j \\ 
T_j/\tau_{j}^{0}\left(Y_{1 j}-\mu_{1}^{0}\right)-\left(T_j-\tau_{j}^{0}\right) \boldsymbol{KD}^{-1} \tilde{\bx}_j \\ 
\{({1-T_j})/({1-\tau_{j}^{0}})-1\}\left\{m_{0 j}^{*}-\operatorname{E}({m}_{0j}^{*})\right\}-\left(T_j-\tau_{j}^{0}\right) \boldsymbol{MD}^{-1} \tilde{\bx}_j \\ 
({1-T_j})\left(Y_{0 j}-\mu_{1}^{0}+\theta^{0}\right)/({1-\tau_{j}^{0}})-\left(T_j-\tau_{j}^{0}\right) \boldsymbol{ND}^{-1} \tilde{\bx}_j\end{array}\right)\,,
\]
where $\boldsymbol{D}$, $\boldsymbol{J}$, $\boldsymbol{K}$, $\boldsymbol{M}$ and $\boldsymbol{N}$ are defined before in Section (\ref{sec::SELR1}). 
We present the limiting distributions for the maximum SEL estimator $\hat{\bfeta}_{\mbox{\tiny SEL2}}=(\hat{\mu}_{1\mbox{\tiny SEL2}}, \hat{\theta}_{\mbox{\tiny SEL2}})^\top$ and SEL ratio statistic for $\theta$ in Theorem \ref{SEL2normal} and Theorem \ref{thm::SEL2chi2app} respectively. 

\begin{theorem}
Under certain regularity conditions and a correctly specified propensity score model, we have $\sqrt{n}\left(\hat{\bfeta}_{\mbox{\tiny SEL2}}-\bfeta^0\right)\stackrel{d}{\rightarrow}\operatorname{MVN}(\boldsymbol{0}, \bV_2)$ as $n\rightarrow\infty$, where $\bV_2=[\boldsymbol{E}^{12}]^\top\boldsymbol{\Omega}_2\boldsymbol{E}^{12}$. 
\label{SEL2normal}
\end{theorem}

\begin{theorem}
Under certain regularity conditions and a correctly specified propensity score model, the scaled profile sample empirical likelihood ratio statistic $-2r_{\mbox{\tiny SEL2}}(\theta)/\delta_2$ at $\theta=\theta^0$ converges in distribution to a $\chi^2_1$-distributed random variable,
where $\delta_2$ is the non-zero eigenvalue of  $\boldsymbol{\Omega}_2^{1/2}\left(\boldsymbol{F}^{11}-\boldsymbol{E}^{11}\right)\boldsymbol{\Omega}_2^{1/2}$.
\label{thm::SEL2chi2app}
\end{theorem}

It follows from Theorem \ref{thm::SEL2chi2app} that a $100(1-\alpha)\%$ confidence interval for the ATE $\theta$ can be constructed as 
 $ \{\theta \mid -2r_{\mbox{\tiny SEL2}}(\theta)/\hat{\delta}_2\leq \chi^2_1(\alpha)\}$,
where $\hat{\delta}_2$ represents the non-zero eigenvalue of $\hat{\boldsymbol{\Omega}}_2^{1/2}(\hat{\boldsymbol{F}}^{11}-\hat{\boldsymbol{E}}^{11})\hat{\boldsymbol{\Omega}}_2^{1/2}$, and $\hat{\boldsymbol{\Omega}}_2$, $\hat{\boldsymbol{E}}^{11}$ and $\hat{\boldsymbol{F}}^{11}$ are the plug-in estimators.

One of the computational challenges for the current problem is the handling of the nuisance parameter $\mu_1$, which prevents a straightforward application of computational techniques for one-sample problems.
Let $\boldsymbol{\phi}=({\boldsymbol{\lambda}}_1, {\boldsymbol{\lambda}}_0, {\mu}_1)^\top$.
For a given $\theta$, the vector of estimators $\hat{\boldsymbol{\phi}}$ solves the equation  
$S_{n2}(\boldsymbol{\phi}, \theta)={\partial \ell_{n2}(\boldsymbol{\phi}, \theta)}/{\partial \boldsymbol{\phi}}=\boldsymbol{0}$, where
\[
\ell_{n2}(\boldsymbol{\phi}, \theta )=-\sum_{j \in \bs_1}\log\left(1+{\boldsymbol{\lambda}}_1^\top\boldsymbol{g}_{1j}^S(\mu_1)\right)-\sum_{j \in \bs_0}\log\left(1+{\boldsymbol{\lambda}}_0^\top\boldsymbol{g}_{0j}^S(\mu_1,\theta)\right)\, .
\]
The Newton-Raphson method is then used to find $\hat{\boldsymbol{\phi}}=(\hat{{\boldsymbol{\lambda}}}_1, \hat{{\boldsymbol{\lambda}}}_0, \hat{\mu}_1)$ for any given $\theta$.

\subsection{Bootstrap Approach for SEL Ratio Confidence Intervals} 

The bootstrap approach based on the current SEL formulation closely resembles the one described in Section \ref{sec::bootstrapSEL}. 
The primary distinction lies in the adjustment required for Step 3 described below.

\begin{description}
    
\item[Step 3] Construct the bootstrap version of the SEL function as
        \begin{equation}
      \ell^{[b]}_{\mbox{\tiny SEL}}(\bp_1, \bp_0) = \sum_{j\in \bs^{[b]}_1}\log(p_{1j})+\sum_{j\in \bs_0^{[b]}}\log(p_{0j})\notag
\end{equation}
    and the bootstrap versions of the constraints in (\ref{eq::SELjointcons2}) as 
    \begin{equation}
        \begin{aligned}
        &\sum_{j\in\bs_1^{[b]}}p_{1j}=1\,, \;\;\;\;\;\;\;\;\;\;\;\;\;\;\;\;\;\;\;\;\;\;\;\;\;\;\;\;\;\;\;\;\;\;\;\;\;\;\;\;
        \sum_{j\in\bs_0^{[b]}}p_{0j}=1\,,\\
        &\sum_{j \in \bs_{1}^{[b]}} p_{1j}\left(\htau_j^{[b]}\right)^{-1}\left(\hm_{1j}^{[b]}-\bar{\hm}_1^{[b]}\right)=0\,,\;\;\;\;\;\;\;\sum_{j \in \bs_{0}^{[b]}} p_{0j}\left(1-\htau_j^{[b]}\right)^{-1}\left(\hm_{0j}^{[b]}-\bar{\hm}_0^{[b]}\right)=0\,,\\
        &\sum_{j \in \bs_{1}^{[b]}} p_{1j}\left(\htau_j^{[b]}\right)^{-1}(Y_{1j}-\mu_1)=0\,,\;\;\;\;\;\;\;\;\;\;\;\;\sum_{j \in \bs_{0}^{[b]}} p_{0j}\left(1-\htau_j^{[b]}\right)^{-1}(Y_{0j}-\mu_1+\theta)=0\,,
        \end{aligned}
        \notag
    \end{equation}
where $\Bar{\hm}_i^{[b]}=\sum_{j\in\bs^{[b]}}\hm_{ij}^{[b]}/n$ for $i=0,1$. Evaluate the profile sample empirical likelihood ratio $r^{[b]}_{\mbox{\tiny SEL2}}(\theta)$ at $\theta=\SELT$ for the $b$-th bootstrap sample, based on the updated objective function and constraints. 
\end{description}

\section{Simulation Studies}
\label{sec::simulation}

In this section, we present results from simulation studies on finite sample performances of point estimators, confidence intervals and powers of tests using our proposed SEL based methods. 

\subsection{Point Estimators and Confidence Intervals}
\label{sec::P&CI}

We conduct simulation studies under the same settings as those in \cite{huang2024pseudoempirical} to assess the performance of the proposed maximum SEL estimators and SEL ratio confidence intervals. 

The sample sizes of the random sample $\bs$ from an infinite population vary across three scenarios: $n=100$ (small sample), $n=200$ (moderate sample), and $n=400$ (large sample).  
Each subject $j$ within the sample $\bs$ possesses covariates denoted as $\bx_j=(x_{j1},x_{j2},x_{j3})^\top$, where $x_{j1}=v_{j1}$, $x_{j2}=v_{j2}+0.2x_{j1}$, and $x_{j3}=v_{j3}+0.3(x_{j1}+x_{j2})$. Here, $v_{j1}\sim \operatorname{N}(0,1)$, $v_{j2}\sim \operatorname{Bernoulli}(0.6)$, and $v_{j3}\sim \operatorname{Exponential}(1)$.
These covariates are instrumental in encapsulating the complexity of real-world scenarios.
The true propensity score model is specified as
\[\tau:\;\;\;\tau_j^0 = \operatorname{expit}(\alpha_0+0.2x_{j1} +0.2x_{j2}-0.5x_{j3})\;\; {\rm for} \; \;  j=1, \dots, n, \] 
where $\alpha_0$ is a parameter controlled by the expected treatment proportion $t$, with $t$ taking values of $0.3$, $0.5$, or $0.7$. 
Different $t$ results in varying proportions of subjects between the treatment and control groups, such as an approximate $3:7$ ratio between the treatment and control subjects for $t=0.3$. 
The treatment indicator $T_j$ follows a Bernoulli distribution with parameter $\tau_j^0$, where $T_j=1$ indicates the experimental treatment and $T_j=0$ represents being assigned to the control group.
The outcome regression models are defined as follows:
\begin{equation}
\begin{aligned}
m_1:\;\;\;Y_{1j}&=4.5+x_{j1}-2x_{j2}+3x_{j3}+a_1\epsilon_j\\
m_0:\;\;\;Y_{0j}&=1+x_{j1}+x_{j2}+2x_{j3}+a_0\epsilon_j\; ,
\end{aligned}  
\notag
\end{equation}
for $j=1, \dots, n$, where $\epsilon_j\sim \operatorname{N}(0,1)$. 
The parameters $a_1$ and $a_0$ are determined to ensure a correlation of $\rho$ between the linear predictor of $\bx_j$ and the corresponding potential outcome, where $\rho$ takes values of $0.3$, $0.5$, or $0.7$. 
A smaller $\rho$ implies higher variability among the potential outcomes $Y_{1j}$'s and $Y_{0j}$'s. 
The true average treatment effect is $\theta^0=2.88$, calculated based on the two outcome regression models.
The observed outcome is either $Y_{1j}$ in the treatment group or $Y_{0j}$ in the control group, given by $Y_j=T_j Y_{1j}+(1-T_j)Y_{0j}$ for $j=1, \dots, n$. 
Ultimately, the accessible data consist of $\{(\bx_j, T_j, Y_j), j\in \bs\}$.
Three scenarios are contemplated for the misspecification in the working models for the propensity score model $\tau$, as well as the outcome regression models $m_1$ and $m_0$:
\begin{itemize}
\item TT: Both the propensity score model and the outcome regression models are correctly specified.
\item TF: The propensity score model is correctly specified, but the two outcome regression models are misspecified by neglecting $x_{j3}$.
\item FT: The outcome regression models are correctly specified, but the propensity score model is misspecified by omitting $x_{j3}$.
\end{itemize}
In summary, a total of 81 distinct settings are investigated by conducting simulations across three different values for each parameter $\rho$, $n$, and $t$, combined with the three misspecification scenarios.

The two SEL approaches proposed in this paper are applied to the simulated data to obtain maximum SEL estimators $\SELO$ and $\SELT$. The SEL ratio confidence intervals are obtained based on the scaled chi-square distribution ($\mathcal{I}_{\mbox{\tiny SELR1}}$ and $\mathcal{I}_{\mbox{\tiny SELR2}}$ ), and through the bootstrap procedure ($\mathcal{I}_{\mbox{\tiny SELR1B}}$ and $\mathcal{I}_{\mbox{\tiny SELR2B}}$). 
The simulation encompasses a total of $n_{sim}=1000$ iterations, and for each iteration where the bootstrap procedure is implemented, the number of bootstrap replications is designated as $B=1000$.
The point estimators are assessed through two measurement metrics: the percentage relative bias ($\%RB$)
and the mean squared error ($MSE$).
The performance of confidence intervals is evaluated using the percentage coverage probability ($\%CP$) and the average length ($AL$). 

Tables \ref{Pt03}, \ref{Pt05}, and \ref{Pt07} display the percentage relative biases and the mean squared errors of the two proposed maximum SEL estimators, when the anticipated treatment proportion $t$ takes a value of $0.3$, $0.5$, and $0.7$, respectively.
The results indicate that both estimators are doubly robust, agreeing with the theoretical results. 
When $t=0.3$ and the outcome regression models are misspecified (``TF"), 
the absolute percentage relative biases for both estimators are over $5\%$ for $n=100$ but decrease as the sample size gets bigger.
It is interesting to notice that the estimator $\SELO$ seems less biased than $\SELT$ for a finite sample when the outcome regression models are misspecified (``TF").

Tables \ref{CIt03}, \ref{CIt05}, and \ref{CIt07} show the percentage coverage probabilities and the average lengths of the confidence intervals for the settings where $t=0.3$, $0.5$, and $0.7$, respectively. 
When the sample size is small, convergence issues arise when computing confidence intervals using the second proposed SEL approach. 
This is likely due to the complexity in the computational procedure introduced by the profiling of $\mu_1$.
In the most extreme case, less than $3\%$ of the simulation samples experience convergence issues, and the problem tends to diminish with increasing sample sizes. 
The reported results are based on the samples without the problem. 
For instance, in the scenario of $(n,t,\rho)=(100, 0.3, 0.3)$ and ``TT", out of 1000 simulation samples, five encounter convergence issues during the computation of $\mathcal{I}_{\mbox{\tiny SELR2}}$, and the simulation results for $\mathcal{I}_{\mbox{\tiny SELR2}}$ in this scenario are derived from the remaining 995 samples.
Overall, the simulation results underscore the following observations:
(1) all four methods work well if both models are correctly specified (``TT"), though the confidence intervals based on bootstrap tend to be wider with higher coverage probabilities compared with the others, especially when the sample size is small; 
(2) the two $\chi^2$-approximation based SEL ratio confidence intervals ($\mathcal{I}_{\mbox{\tiny SELR1}}$ and $\mathcal{I}_{\mbox{\tiny SELR2}}$) seem to be doubly robust to the misspecification of either the propensity score model or the outcome regression models, while $\mathcal{I}_{\mbox{\tiny SELR2}}$ is usually narrower than $\mathcal{I}_{\mbox{\tiny SELR1}}$ with comparable coverage probabilities when the sample size is small; 
(3) when $t$ is $0.3$, the propensity score model is misspecified, and $n$ takes the value of $100$ or $200$, the coverage probabilities of both $\mathcal{I}_{\mbox{\tiny SELR1}}$ and $\mathcal{I}_{\mbox{\tiny SELR2}}$ are slightly lower than the nominal level of $95\%$, but these two confidence intervals work well in all the other cases;
and (4) the bootstrap-calibrated confidence intervals ($\mathcal{I}_{\mbox{\tiny SELR1B}}$ and $\mathcal{I}_{\mbox{\tiny SELR2B}}$) are doubly robust, though their coverage probabilities occasionally hover around $97\%$, indicating a conservative nature, particularly when the sample size is small.

\subsection{Power of Statistical Significance Tests}
It is often of interest to test the significance of a causal effect. That is, we wish to test the hypotheses $H_0: \theta=0$ versus $H_a: \theta \ne 0$. 
This can be done by using confidence intervals. The null hypothesis $H_0$ is rejected at a significance level of $5\%$ if the value of $0$ is not covered by the $95\%$ confidence interval $\mathcal{I}$ for the parameter $\theta$. 

\begin{figure}
\begin{tabular}{ccc}
  \includegraphics[width=0.33\textwidth]{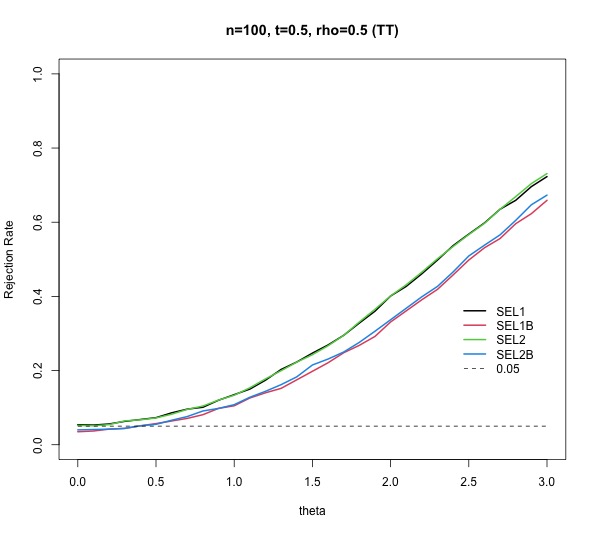} & \includegraphics[width=0.33\textwidth]{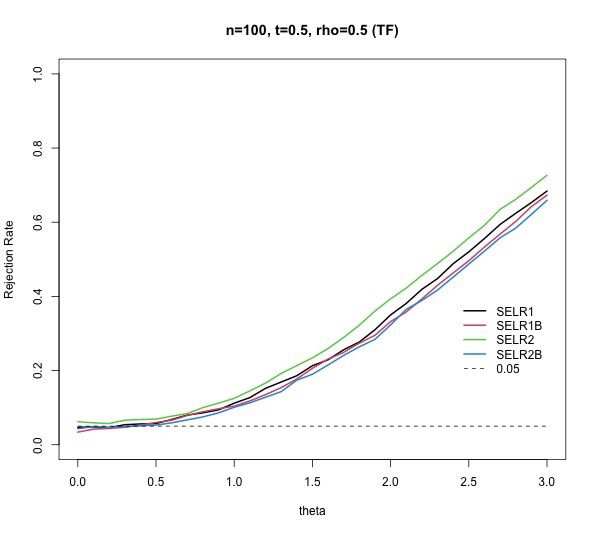}&
  \includegraphics[width=0.33\textwidth]{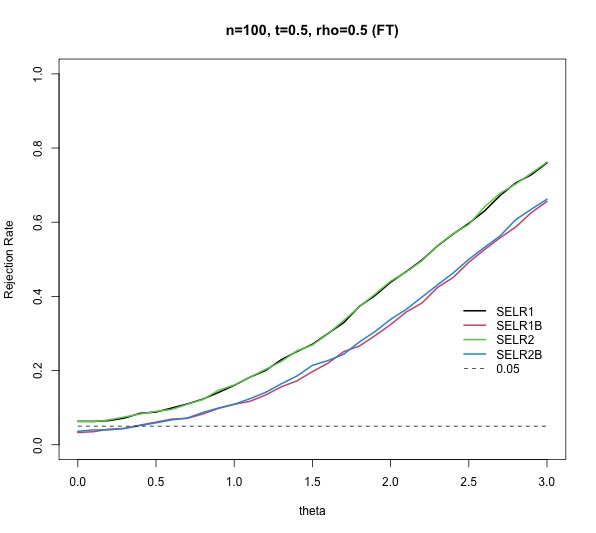}\\
  \includegraphics[width=0.33\textwidth]{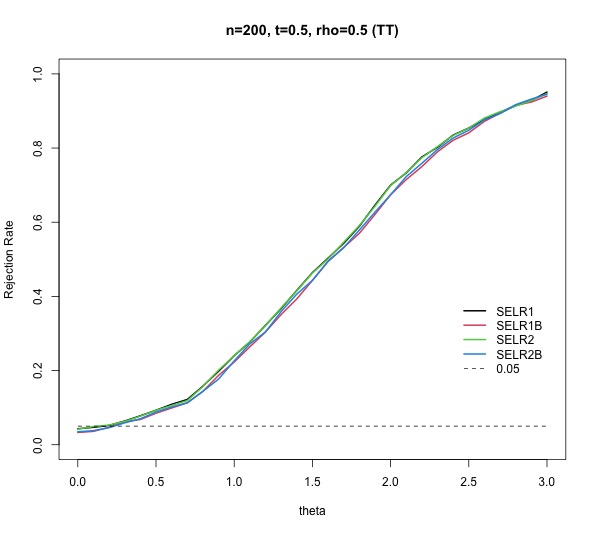} & \includegraphics[width=0.33\textwidth]{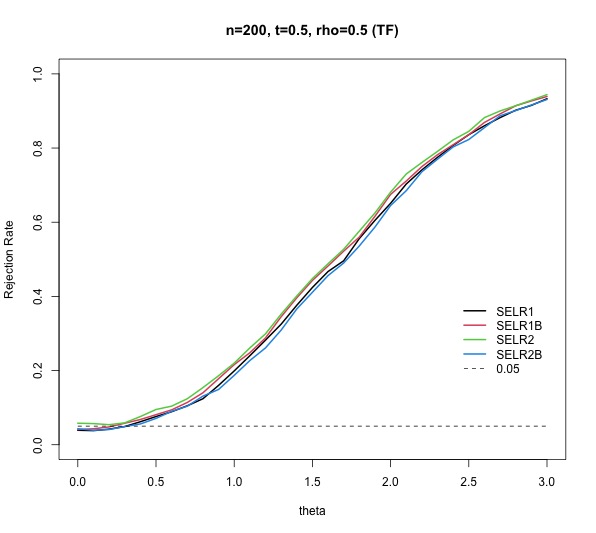}&
  \includegraphics[width=0.33\textwidth]{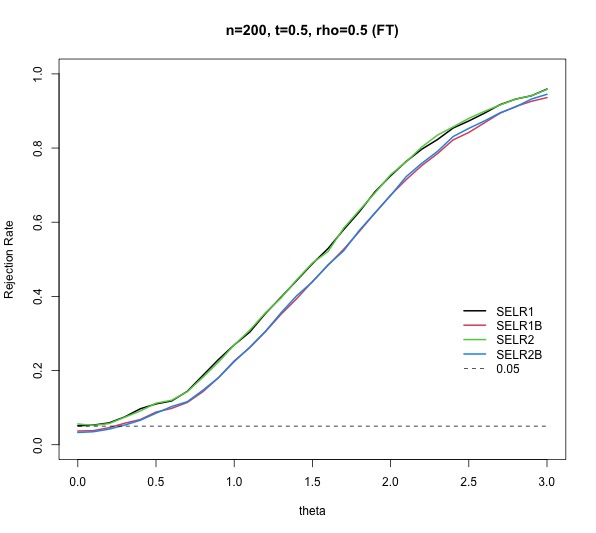}\\
  \includegraphics[width=0.33\textwidth]{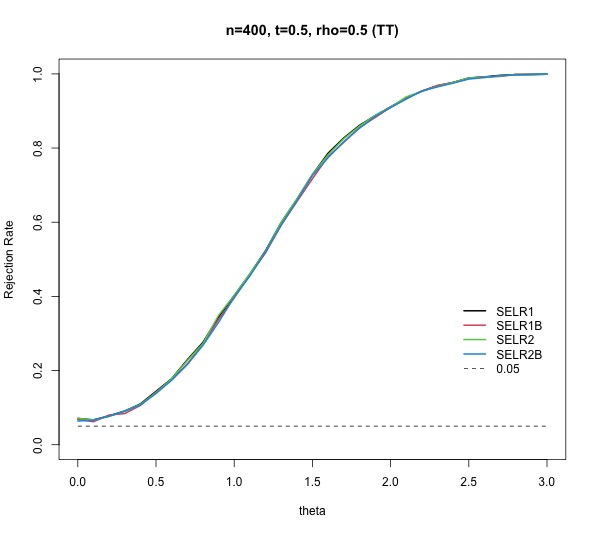} & \includegraphics[width=0.33\textwidth]{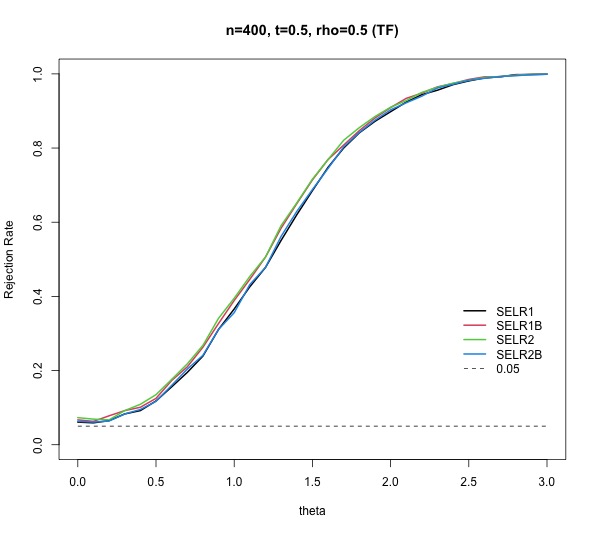}&
  \includegraphics[width=0.33\textwidth]{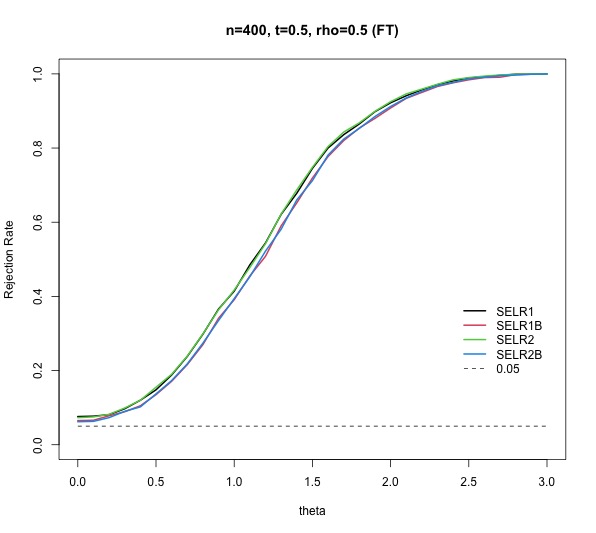}
\end{tabular}
\caption{Power functions of tests when $t=0.5$ and $\rho=0.5$.}
\label{rejectiont05rho05}
\end{figure}

A good statistical test should have high power and the type I error close to the nominal significance level. 
In simulations, 
the rejection rate, 
\[
\text{Rejection Rate}=\frac{1}{n_{sim}}\sum_{s=1}^{n_{sim}}I\left(0\notin\mathcal{I}^{(s)}\right)\, ,
\]
is the empirical type I error probability if the true ATE $\theta$ is $0$, and it is the empirical power otherwise. 
We want to evaluate the rejection rate at a sequence of different true values for $\theta$ to see how the proposed SEL ratio tests perform. 
To do so, we specify the two outcome regression models as follows:
\begin{equation}
\begin{aligned}
m_1:\;\;\;Y_{1j}&=\theta^0+4.5+x_{j1}-2x_{j2}+3x_{j3}+a_1\epsilon_j\\
m_0:\;\;\;Y_{0j}&=3.88+x_{j1}+x_{j2}+2x_{j3}+a_0\epsilon_j\; ,
\end{aligned}  
\notag
\end{equation}
where $\theta^0$ is the assigned true value of the ATE and varies from $0$ to $3$. 
The other settings are the same as those in Section \ref{sec::P&CI}. 
Four different types of SEL ratio confidence intervals are calculated from each simulation and used for testing the null hypothesis $H_0: \theta=0$. 
The curve of the power function is produced by plotting rejection rates over a range of true values for $\theta$ when using each type of confidence interval. 
Figure \ref{rejectiont05rho05} shows the curves for different combinations of sample sizes and model misspecifications when $t=0.5$ and $\rho=0.5$.
The plots highlight that: 
(1) the curves produced by using the $\chi^2$-approximation consistently exhibit higher power compared to the other two curves based on bootstrap procedures for all the cases when $n=100$ and all the cases when the PS model is misspecified; 
(2) as the sample size increases, the curves for all approaches tend to coincide, indicating similar performance in large samples;
(3) notably, the curves of all approaches increase as the true value of the ATE diverges further from $0$; especially when the sample size is large, they move close to $1$ quickly, implying all four approaches work well.

\section{Conclusions and Additional Remarks}
\label{sec::discussion}
We propose two different approaches using the sample empirical likelihood for statistical inference of the average treatment effect. 
Theoretical properties of the two approaches are studied, and asymptotic results are established. 
Simulation studies are also conducted to confirm the numerical performance.
Our proposed point estimators are doubly robust against misspecifications of the propensity score model and the outcome regression model, and the confounders are balanced explicitly based on these two methods.
Our proposed SEL ratio confidence intervals posses desirable properties such as range-respecting and transformation-invariant, which are attractive when the true ATE is in a specific range; for example, the outcome $Y$ is a binary variable, i.e., the ATE is in the range of $(-1,1)$. 

The flexibility of our frameworks allows for the incorporation of more appropriate constraints, enabling the derivation of multiply robust estimators \citep{Han2013} or more efficient estimators. 
For instance, various models can be employed for propensity score modelling and outcome regression modelling, all of which can be introduced into the constraints. 
As a result, the resulting estimator becomes multiply robust, ensuring consistency if any one of the multiple models is correctly specified.
On the other hand, suitable information from external datasets may also be incorporated by adding additional constraints, facilitating the construction of more efficient estimators.

\section{Appendix}

\subsection{Regularity Conditions}

The regularity conditions are the same as those listed in Section 7.1 in \cite{huang2024pseudoempirical}. 
Let $\tau(\bx; \balpha)$ be specified by the propensity score model, and  $m_1(\bx, \bbeta_1)$ and $m_0(\bx, \bbeta_0)$ be the mean functions of outcome regression models for potential outcomes $Y_1$ and $Y_0$ given $\bx$.
The probability limit of the estimator $\hat{\bbeta}_i$ under the assumed working model using the observed sample data is denoted as $\bbeta_i^*$ for $i=0,1$.

\begin{itemize}
    \item [{\bf R1}] The treatment indicator $T$ satisfies $E(T)=t\in (0,1)$.

    \item [{\bf R2}] The population satisfies $\operatorname{E}({Y}_{1}^2)<\infty$, $\operatorname{E}({Y}_{0}^2)<\infty$, and $\operatorname{E}(\|\bx\|^2)<\infty$, where $\|\cdot \|$ denotes the $l_2$-norm. 

    \item [{\bf R3}] The population and the mean functions satisfy $\operatorname{E}\{m_1^2(\bx,\bbeta_1^*)\}<\infty$ and $\operatorname{E}\{m_0^2(\bx,$\\$\bbeta_0^*)\}<\infty$.

    \item [{\bf R4}] For each $\bx$, $\partial m_i(\bx, \bbeta_i)/\partial\bbeta_i$ is continuous in $\bbeta_i$ and $\|\partial m_i(\bx, \bbeta_i)/\partial\bbeta_i \|\leq h_i(\bx, \bbeta_i)$ for $\bbeta_i$ in the neighborhood of $\bbeta_i^*$, and $\operatorname{E}\{h_i(\bx, \bbeta_i^*)\}<\infty$, for $i=0,1$. 

     \item [{\bf R5}] For each $\bx$, $\partial^2 m_i(\bx, \bbeta_i)/\partial\bbeta_i\partial\bbeta_i^\top$ is continuous in $\bbeta_i$ and $\operatorname{max}_{j,l}|\partial^2 m_i(\bx, \bbeta_i)/\partial\bbeta_{ij}$\\$\partial\bbeta_{il}|\leq k_i(\bx, \bbeta_i)$ for $\bbeta_i$ in the neighborhood of $\bbeta_i^*$, and $\operatorname{E}\{k_i(\bx, \bbeta_i^*)\}<\infty$, for $i=0,1$, where $|\cdot|$ denotes the absolute value.

     \item[{\bf R6}] There exist $c_1$ and $c_2$ such that $0<c_1\leq \tau_j^0\leq c_2<1$ for all units $j$, where $\tau_j^0=\tau(\bx_j;\balpha^0)$ is the propensity score under the true propensity score model and $\balpha^0$ is the true value for the vector of propensity score model parameters $\balpha$.
     
\end{itemize}

\subsection{Proof of Therorem \ref{thm::SEL1mu1}}
\label{sec::ProofDRmut}
\begin{proof}
We justify the double robustness of $\hat{\mu}_{1\mbox{\tiny SEL1}}$ by demonstrating the consistency when the PS model or the outcome regression model is correctly specified. 

\noindent
{\em (1) The outcome regression model is correctly specified.}

The model-calibration constraint in (\ref{eq::jcal}) for the treated sample $\bs_1$ guarantees that $\hat{\mu}_{1\mbox{\tiny SEL1}} = \sum_{j\in \bs_1}\hat{p}_{1j} Y_{1j}=\sum_{j\in \bs_1}\hat{p}_{1j}(Y_{1j}-\hm_{1j})+\bar{\hm}_1$.
Under regularity conditions {\bf R1}-{\bf R4}, plugging in the expression for $\hat{p}_{1j}$ yields 
\begin{equation}
    \begin{aligned}
    \hat{\mu}_{1\mbox{\tiny SEL1}}&=\frac{1}{n_{1}} \sum_{j \in \bs} T_j \frac{Y_{1 j}-\hm_{1 j}}{1+\hat{\blambda}_1^{\top} \hat{\boldsymbol{v}}_{1j}}+\frac{1}{n} \sum_{j \in \bs} \hm_{1 j} \\ 
    &=\frac{1}{n_{1}} \sum_{j \in \bs} T_j \frac{Y_{1j}-m_{1j}^{0}}{1+{\blambda_{1}^{*}}^\top \boldsymbol{v}_{1j}\left(\balpha^{*}, \bbeta_{1}^0\right)}+\frac{1}{n} \sum_{j \in \bs} m_{1 j}^{0}+o_p(1)
    \end{aligned}
    \notag
\end{equation}
where $\balpha^{*}$ and $\bbeta_{1}^0$ are the limiting values for $\hat{\balpha}$ and $\hat{\bbeta}_1$, and $\blambda^{*}_{1}$ and $m_{1 j}^{0}$ are the corresponding limiting values of $\hat{\blambda}_1$ and $\hm_{1j}$.
It converges in probability to 
\[
\frac{1}{t} \operatorname{E}\left(\frac{T_j\left(Y_{1j}-m_{1 j}^{0}\right)}{1+\blambda_{1*}^{\top} \boldsymbol{v}_{1j}\left(\balpha_{*}, \bbeta_{1}^0\right)}\right)+\mu_{1}^{0}
\]
under regularity condition {\bf R1}. 
The expectation term takes a value of zero under a correctly specified outcome regression model, 
therefore $\hat{\mu}_{1\mbox{\tiny SEL1}}$ converges to $\mu_1^0$ in probability in this case.

\noindent
{\em (2) The PS model is correctly specified.}

We know that 
\begin{equation}
    \begin{aligned}
        \hat{\mu}_{1\mbox{\tiny SEL1}} = \sum_{j\in \bs_1}\hat{p}_{1j} Y_{1j}=
    \sum_{j\in \bs_1}\frac{\Bar{\htau}}{\htau_j}\hat{q}_{1j}Y_{1j}
    =\frac{\bar{\htau}}{n_1}\sum_{j\in\bs_1}\frac{Y_{1j}/\htau_j}{1+\hat{\blambda}_{1q}^\top\hat{\boldsymbol{v}}_{1j}/\htau_j}\, ,
    \end{aligned}
    \notag
\end{equation}
where $\hat{\blambda}_{1q}$ is the solution of equation (\ref{eq::SELLagequation}).
From equation (\ref{eq::SELLagequation}), we obtain the inequality
\[
\left \|\frac{1}{n}\sum_{j\in\bs}T_j\hat{\boldsymbol{v}}_{1j}/\htau_j\right \| \geq \left \|\hat{\blambda}_{1q}\right \|\frac{1}{1+\left \|\hat{\blambda}_{1q}\right \|\max_{j} \|\hat{\boldsymbol{v}}_{1j}/\htau_j \|} \left \|\frac{1}{n}\sum_{j\in\bs}\frac{T_j}{\htau_j}\frac{\hat{\boldsymbol{v}}_{1j}\hat{\boldsymbol{v}}_{1j}^\top}{\htau_j}\right \|\, ,
\]
which reveals that $\hat{\blambda}_{1q}=O_p(n^{-1/2})$ under regularity conditions {\bf R1} and {\bf R3}-{\bf R6}, and further $\hat{\blambda}^\top_{1q}\hat{\boldsymbol{v}}_{1j}/\htau_j=o_p(1)$ uniformly for all $j\in\bs_1$.
Therefore, under regularity conditions {\bf R1}-{\bf R6}, taking the Taylor expansion of $\hat{\mu}_{1\mbox{\tiny SEL1}}$ at  $\hat{\blambda}^\top_{1q}\hat{\boldsymbol{v}}_{1j}/\htau_j=0$ provides 
\[
\hat{\mu}_{1\mbox{\tiny SEL1}} =\frac{\sum_{j\in\bs}\htau_j}{n_1}\left(\frac{1}{n}\sum_{j\in \bs_1}\frac{Y_{1j}}{\htau_j}+O_p(n^{-1/2})\right)=\mu_1^0+O_p(n^{-1/2})\, ,
\]
where we use the fact that $\sum_{j\in\bs}\htau_j/n_1=1+O_p(n^{-1/2})$. 
Thus, $\hat{\mu}_{1\mbox{\tiny SEL1}}$ is a valid estimator when the PS model is correctly specified.
\end{proof}

\subsection{Proof of Theorem \ref{thm::SEL1normalPoint}}
\label{sec::ProofSEL1Point}
\begin{proof} Recall the maximum SEL estimators $\hat{\mu}_{\mbox{\tiny SEL1}}$ and $\hat{\theta}_{\mbox{\tiny SEL1}}$ maximize the profile SEL function $\ell(\hat{\bq}_1(\mu_1, \theta), \hat{\bq}_0(\mu_1, \theta))$. Let 
\begin{eqnarray*}
\ell_n(\mu_1, \theta) &=& \ell(\hat{\bq}_1(\mu_1, \theta), \hat{\bq}_0(\mu_1, \theta))  \\
&=& - \sum_{j \in \bs_1}\log\left(1+\hat{\blambda}_1^\top\boldsymbol{g}_{1j}(\mu_1)\right)-\sum_{j \in \bs_0}\log\left(1+\hat{\blambda}_0^\top\boldsymbol{g}_{0j}(\mu_1,\theta)\right)\, ,
\end{eqnarray*}
where constant terms are omitted, and the Lagrange multipliers $\hat{\blambda}_1$ and $\hat{\blambda}_0$ are solutions of 
equations (\ref{eq::SELJointLagEqua}) with the following expressions 
\[ \sum_{j \in \bs_1}\frac{\boldsymbol{g}_{1j}(\mu_1)}{1+\blambda_1^\top\boldsymbol{g}_{1j}(\mu_1)}=\boldsymbol{0} \;\;\;{\rm and }\;\;\;\sum_{j \in \bs_0}\frac{\boldsymbol{g}_{0j}(\mu_1, \theta)}{1+\blambda_0^\top\boldsymbol{g}_{0j}(\mu_1, \theta)}=\boldsymbol{0}\,. 
\]

Define $\bfeta=(\blambda_1^\top, \blambda_0^\top, \mu_1, \theta)^\top$ and a function based on $(\blambda_1, \blambda_0, \mu_1, \theta)$ that 
\[
\ell_n(\blambda_1, \blambda_0, \mu_1, \theta )=-\sum_{j \in \bs_1}\log\left(1+{\blambda}_1^\top\boldsymbol{g}_{1j}(\mu_1)\right)-\sum_{j \in \bs_0}\log\left(1+{\blambda}_0^\top\boldsymbol{g}_{0j}(\mu_1,\theta)\right)\, .
\]
The vector of estimators $\hat{\bfeta}=(\hat{\blambda}_{1\mathrm{\mbox{\tiny SEL1}}}^\top, \hat{\blambda}_{0\mathrm{\mbox{\tiny SEL1}}}^\top, \hat{\mu}_{1\mathrm{\mbox{\tiny SEL1}}},\SELO)^\top$ satisfies that
\[
S_n(\bfeta)=\frac{\partial \ell_n(\bfeta)}{\partial \bfeta^\top}=\boldsymbol{0}\, .
\]
Note that $\hat{\blambda}_{i\mathrm{\mbox{\tiny SEL1}}}=\hat{\blambda}_{i}(\hat{\mu}_{1\mathrm{\mbox{\tiny SEL1}}}, \SELO)$, where $\hat{\blambda}_{i}(\mu_1, \theta)$ denotes the solution of equations (\ref{eq::SELJointLagEqua}) for a given $(\mu_1, \theta)$ for $i=0,1$.
Since $(\hat{\mu}_{1\mathrm{\mbox{\tiny SEL1}}}, \hat{\theta}_{\mathrm{SEL1}})=(\mu_1^0, \theta^0)+O_p(n^{-1/2})$ when the PS model is correctly specified, under regularity conditions {\bf R1}-{\bf R6}, the property still holds that $\hat{\blambda}_{i\mathrm{\mbox{\tiny SEL1}}}=O_p(n^{-1/2})$ for $i=0,1$.  

Let $\bfeta^0=(\boldsymbol{0}^\top,\boldsymbol{0}^\top,\mu_1^0, \theta^0)^\top$.
Under regularity conditions {\bf R1}-{\bf R6}, applying the Taylor expansion of $S_n(\hat{\bfeta})$ at $\bfeta=\bfeta^0$ leads to
\[
\frac{1}{n}S_n(\bfeta^0)+\frac{1}{n}\frac{\partial S_n(\bfeta^0)}{\partial \bfeta^\top}(\hat{\bfeta}-\bfeta^0)+o_p(n^{-1/2})=\boldsymbol{0}\, .
\] 
This implies
\[
\hat{\bfeta}-\bfeta^0=-\frac{1}{n}\bA^{-1}S_n(\bfeta^0)+o_p(n^{-1/2})\, ,
\]
where $\bA$ is the limit of $n^{-1}\partial S_n(\bfeta^0)/\partial \bfeta^\top$ and its expression is given in Section \ref{sec::SELR1}. 
On the other hand, denote 
\[
S_{n}^{\ast}(\bfeta^0)=-\sum_{j\in\bs}\begin{pmatrix}
T_j\boldsymbol{g}_{1j}(\mu_1^0)\\
(1-T_j)\boldsymbol{g}_{0j}(\mu_1^0, \theta^0)
\end{pmatrix} 
\]
and we have 
\[
-\frac{1}{n}S_n(\bfeta^0)=
\frac{1}{n}\sum_{j \in \bs}
\begin{pmatrix}
T_j\boldsymbol{g}_{1j}(\mu_1^0)\\
(1-T_j)\boldsymbol{g}_{0j}(\mu_1^0, \theta^0)\\0\\0
\end{pmatrix}
=\begin{pmatrix}
-\frac{1}{n}S_{n}^{\ast}(\bfeta^0)\\
\boldsymbol{0}
\end{pmatrix}\,.
\]

We apply estimating equation techniques to rewrite $-S_{n}^{\ast}(\bfeta^0)/n$ in the form of a sample mean of independent and identically distributed random variables to obtain its limiting distribution, which is also influenced by the estimation procedures of the propensity score and outcome regressions.
The derivation of the second element of $-S_{n}^{\ast}(\bfeta^0)/n$ is shown below as an example.
Applying Taylor expansion for the second element of $-S_{n}^{\ast}(\bfeta^0)/n$ at $\hat{\bbeta}_1=\bbeta_1^*$ yields that
\[
\frac{1}{n}\sum_{j\in\bs} T_j \frac{1}{\hat{\tau}_j}\left(\hat{m}_{1j}-\Bar{\hat{m}}_1\right)=\frac{1}{n}\sum_{j\in\bs} T_j \frac{1}{\hat{\tau}_j}\left({m}_{1j}^*-\Bar{{m}}_1^*\right)+o_p(n^{-1/2})\, .
\]
Let a vector of parameters $\boldsymbol{\Psi}=(\gamma, \Bar{m}_1, \balpha^\top)^\top$ and a set of estimating functions 
\[
\boldsymbol{U}(\boldsymbol{\Psi})=
\begin{pmatrix}
\frac{1}{n}\sum_{j\in\bs}\frac{T_j}{\tau_j}\left({m}_{1j}^*-\Bar{{m}}_1\right)-\gamma \\
\frac{1}{n}\sum_{j\in\bs} {m}_{1j}^*-\Bar{{m}}_1\\
\frac{1}{n}\sum_{j\in\bs}\tilde{\bx}_j(T_j-\tau_j)
\end{pmatrix}\, .
\]
The solution to $\boldsymbol{U}(\boldsymbol{\Psi})=\boldsymbol{0}$, i.e., $\hat{\boldsymbol{\Psi}}=(n^{-1}\sum_{j\in\bs}{T_j}({m}_{1j}^*-\Bar{{m}}_1^*)/\hat{\tau}_j, \Bar{{m}}_1^*, \hat{\balpha}^\top)^\top$, is a consistent estimator of $\boldsymbol{\Psi}^0=(0, \operatorname{E}({m}_{1j}^*), \{\balpha^0\}^\top)^\top$.
Moreover, it follows that
\[
\hat{\boldsymbol{\Psi}}-\boldsymbol{\Psi}^0=-\left[\operatorname{E}\{\boldsymbol{H}(\boldsymbol{\Psi}^0)\}\right]^{-1}\boldsymbol{U}(\boldsymbol{\Psi}^0)+o_p(n^{-1/2})\, 
\]
where $\boldsymbol{H}(\boldsymbol{\Psi})=\partial \boldsymbol{U}(\boldsymbol{\Psi})/\partial \boldsymbol{\Psi}$. Therefore, we have 
\[\scalemath{0.94}{
\frac{1}{n}\sum_{j\in\bs}  \frac{T_j}{\hat{\tau}_j}\left(\hat{m}_{1j}-\Bar{\hat{m}}_1\right)=\frac{1}{n}\sum_{j\in\bs} \left\{\left(\frac{T_j}{\tau_{j}^{0}}-1\right)\left(m_{1j}^{*}-\operatorname{E}(m_{1j}^*)\right)+\left(T_j-\tau_{j}^{0}\right) \boldsymbol{J}\boldsymbol{D}^{-1} \tilde{\bx}_j\right\}+o_p(n^{-1/2})}\, .
\]
Repeating the above method for every element of $-S_{n}^{\ast}(\bfeta^0)/n$ leads to 
\begin{equation}
    -\frac{1}{n}S_{n}^{\ast}(\bfeta^0)=\frac{1}{n}\sum_{j \in \bs}\boldsymbol{b}_j+o_p(n^{-1/2})\,,
    \label{eq::S to bj}
\end{equation}
where $\boldsymbol{b}_j$ is given in equation (\ref{eq::bj}).

This expression (\ref{eq::S to bj}) implies that $-\sqrt{n}^{-1}S_{n}^{*}(\bfeta^0)\stackrel{d}{\rightarrow}\operatorname{MVN}(\boldsymbol{0}, \boldsymbol{\Omega}_1)$ under regularity conditions {\bf R2}, {\bf R3} and {\bf R6}, and furthermore, yields that
\[
\sqrt{n}\left(\hat{\bfeta}_{1\mbox{\tiny SEL1}}-\bfeta_1^0\right)\stackrel{d}{\rightarrow}\operatorname{MVN}(\boldsymbol{0}, \boldsymbol{V}_1)\; {\rm as} \; n\rightarrow\infty\,,  
\]
where $\boldsymbol{V}_1=\boldsymbol{A}^{21}\boldsymbol{\Omega}_1\boldsymbol{A}^{12}$, and $\boldsymbol{\Omega}_1$ represents the asymptotic variance-covariance matrix of the vector  
\[
\frac{1}{\sqrt{n}}\sum_{j \in \bs}\begin{pmatrix}
T_j\boldsymbol{g}_{1j}(\mu_1^0)\\
(1-T_j)\boldsymbol{g}_{0j}(\mu_1^0, \theta^0)
\end{pmatrix}\;. 
\]
Note that $\boldsymbol{\Omega}_1$ is also the variance-covariance matrix of the random variable $\boldsymbol{b}$, whose realization is $\boldsymbol{b}_j$.
\end{proof}

\subsection{Proof of Theorem \ref{thm::SEL1chi2}}
\label{sec::ProofSEL1chi2}
\begin{proof}
The sample empirical likelihood ratio statistic at $\theta=\theta^0$ is provided by
\begin{equation} \scalemath{0.95}{   
    \begin{aligned}
     r_{\mbox{\tiny SEL1}}(\theta^0)
     =&\ell_{\mbox{\tiny SEL}}\left(\hat{\bq}_1\left(\hat{\mu}_1\left(\theta^0\right), \theta^0\right), \hat{\bq}_0\left( \hat{\mu}_1\left(\theta^0\right), \theta^0\right)\right)-\ell_{\mbox{\tiny SEL}}\left(\hat{\bq}_1\left(\mutSELO, \SELO\right), \hat{\bq}_0\left(\mutSELO, \SELO\right)\right)\\
     =&\ell_n\left(\Tilde{\blambda}_1, \Tilde{\blambda}_0, \Tilde{\mu}_1, \theta^0\right)-\ell_n\left(\hat{\blambda}_{1\mathrm{\mbox{\tiny SEL1}}}, \hat{\blambda}_{0\mathrm{\mbox{\tiny SEL1}}}, \hat{\mu}_{1\mathrm{\mbox{\tiny SEL1}}},\SELO\right)\\
     =&\left [ \ell_n\left(\Tilde{\blambda}_1, \Tilde{\blambda}_0, \Tilde{\mu}_1, \theta^0\right)-\ell_n\left(\boldsymbol{0},\boldsymbol{0},\mu_1^0, \theta^0\right) \right]\\
     &-\left[ \ell_n\left(\hat{\blambda}_{1\mathrm{\mbox{\tiny SEL1}}}, \hat{\blambda}_{0\mathrm{\mbox{\tiny SEL1}}}, \hat{\mu}_{1\mathrm{\mbox{\tiny SEL1}}},\SELO\right)-\ell_n\left(\boldsymbol{0},\boldsymbol{0},\mu_1^0, \theta^0\right)\right]\, ,
    \end{aligned}}
    \notag
\end{equation}
where $\Tilde{\mu}_1=\hat{\mu}_1(\theta^0)$ and $\Tilde{\blambda}_i=\hat{\blambda}_i(\Tilde{\mu}_1,\theta^0)$. 
Applying the Taylor expansion yields that 
\begin{equation}
\begin{aligned}
&\ell_n\left(\hat{\blambda}_{1\mathrm{\mbox{\tiny SEL1}}}, \hat{\blambda}_{0\mathrm{\mbox{\tiny SEL1}}}, \hat{\mu}_{1\mathrm{\mbox{\tiny SEL1}}},\SELO\right)-\ell_n\left(\boldsymbol{0},\boldsymbol{0},\mu_1^0, \theta^0\right)\\
=& \frac{\partial \ell_n(\bfeta^0)}{\partial \bfeta}\left(\hat{\bfeta}-\bfeta^0\right)+\frac{1}{2}\left(\hat{\bfeta}-\bfeta^0\right)^\top\frac{\partial^2 \ell_n(\bfeta^0)}{\partial\bfeta^\top\partial\bfeta}\left(\hat{\bfeta}-\bfeta^0\right)+o_p(1)\\
=&-\frac{1}{2n}\left[ S_n\left(\bfeta^0\right)\right ]^\top\bA^{-1}S_n\left(\bfeta^0\right)+o_p(1)\\
=&-\frac{1}{2n}\left[ S_{n}^{*}\left(\bfeta^0\right ]\right)^\top\bA^{11}S_{n}^{\ast}\left(\bfeta^0\right)+o_p(1)
\end{aligned}    
\notag
\end{equation}

Next, we use the techniques in the proof of Theorem \ref{thm::SEL1normalPoint} to find the asymptotic properties for $\Tilde{\boldsymbol{\phi}}=(\Tilde{\blambda}_1^\top, \Tilde{\blambda}_0^\top, \Tilde{\mu}_1)^\top$. Fix $\theta$ at $\theta^0$ and consider $\Tilde{\mu}_1=\mu_1^0+O_p(n^{-1/2})$. We still have that $\Tilde{\blambda}_i=O_p(n^{-1/2})$ when the PS model is correctly specified, for $i=0,1$. 
Let $\boldsymbol{\phi}=({\blambda}_1^\top, {\blambda}_0^\top, {\mu}_1)^\top$, $\boldsymbol{\phi}^0=(\boldsymbol{0}^\top, \boldsymbol{0}^\top, {\mu}_1^0)^\top$, and $S_n(\boldsymbol{\phi}, \theta^0)={\partial \ell_n(\boldsymbol{\phi}^\top, \theta^0)}/{\partial \boldsymbol{\phi}^\top}$. 
It follows that $S_n(\tilde{\boldsymbol{\phi}}, \theta^0)=\boldsymbol{0}$. 
Applying the Taylor expansion leads that
\[
\tilde{\boldsymbol{\phi}}-\boldsymbol{\phi}^0=-\frac{1}{n}\boldsymbol{B}^{-1}S_n\left(\boldsymbol{\phi}^0, \theta^0\right)+o_p(n^{-1/2})\, ,
\]
where $\boldsymbol{B}$ is defined in Section \ref{sec::SELR1} and $S_n(\boldsymbol{\phi}^0, \theta^0)=(S_{n}^{\ast}(\bfeta^0)^\top, 0)^\top$. 
By applying the Taylor expansion, we get
\begin{equation}
\begin{aligned}
&\ell_n\left(\Tilde{\blambda}_1, \Tilde{\blambda}_0, \Tilde{\mu}_1, \theta^0\right)-\ell_n\left(\boldsymbol{0},\boldsymbol{0},\mu_1^0, \theta^0\right)\\
=&\frac{\partial \ell_n(\{\boldsymbol{\phi}^0\}^\top,\theta^0)}{\partial \boldsymbol{\phi}} \left(\tilde{\boldsymbol{\phi}}-\boldsymbol{\phi}^0\right)+\frac{1}{2}\left(\tilde{\boldsymbol{\phi}}-\boldsymbol{\phi}^0\right)^\top\frac{\partial^2 \ell_n(\{\boldsymbol{\phi}^0\}^\top, \theta^0)}{\partial\boldsymbol{\phi}^\top\partial\boldsymbol{\phi}}\left(\tilde{\boldsymbol{\phi}}-\boldsymbol{\phi}^0\right)+o_p(1)\\
=&-\frac{1}{2n}\left[S_n\left(\boldsymbol{\phi}^0,\theta^0\right)\right]^\top\boldsymbol{B}^{-1}S_n\left(\boldsymbol{\phi}^0, \theta^0\right)+o_p(1)\\
=&-\frac{1}{2n}\left[S_{n}^{\ast}\left(\bfeta^0\right)\right]^\top\boldsymbol{B}^{11}S_{n}^{\ast}\left(\bfeta^0\right)+o_p(1)
\end{aligned}    
\notag
\end{equation}

Finally, the sample empirical likelihood ratio statistic at $\theta=\theta^0$ has the expression 
\[
r_{\mbox{\tiny SEL1}}(\theta^0)=\frac{1}{2n}\left[ S_{n}^{\ast}\left(\bfeta^0\right)\right]^\top\left(\bA^{11}-\boldsymbol{B}^{11}\right)S_{n}^{\ast}\left(\bfeta^0\right)+o_p(1)\,.
\]
The above equation implies that 
\[
-2r_{\mbox{\tiny SEL1}}(\theta^0)\stackrel{d}{\rightarrow}\boldsymbol{Q}_1^\top\boldsymbol{\Omega}_1^{1/2}\left(\boldsymbol{B}^{11}-\bA^{11}\right)\boldsymbol{\Omega}_1^{1/2}\boldsymbol{Q}_1\, ,
\]
where $\boldsymbol{Q}_1\sim\operatorname{MVN}(\boldsymbol{0}, \boldsymbol{I})$, and $\boldsymbol{I}$ is a $6\times 6$ dimensional identity matrix. 
In the end, we conclude that $-2r_{\mbox{\tiny SEL1}}(\theta^0)\stackrel{d}{\rightarrow}\delta_1\chi^2_1$, where $\delta_1$ is the non-zero eigenvalue of $\boldsymbol{\Omega}_1^{1/2}\left(\boldsymbol{B}^{11}-\bA^{11}\right)\boldsymbol{\Omega}_1^{1/2}$.
\end{proof}

\renewcommand{\bibsection}{\section*{\normalsize REFERENCES}}
\bibliographystyle{chicago}
\bibliography{uw-ethesis}

\newpage
\begin{table}[pt]
    \caption{$\%RB$ and $MSE(\times100)$ for Point Estimators when $t=0.3$}
    \centering
    \begin{tabular}{cclrrrrrr}
    \hline\rule{0pt}{2.1ex}
    \multirow{2}{*}{n}& \multirow{2}{*}{Scenario}&\multirow{2}{*}{Estimator}&\multicolumn{2}{c}{$\rho=0.3$}&\multicolumn{2}{c}{$\rho=0.5$}&\multicolumn{2}{c}{$\rho=0.7$}\\
    \cline{4-9}\rule{0pt}{2.1ex}
    & & &$\% RB$ & $MSE$ &$\% RB$ & $MSE$& $\% RB$ & $MSE$\\
    \hline\rule{0pt}{2.4ex}
        \multirow{6}{*}{100}&\multirow{2}{*}{TT}&$\SELO$&-1.2  & 695.0   & -1.0    & 206.6 & -0.8 & 74.3 \\
    & &$\SELT$&-1.4  & 688.7 & -1.0    & 206.9 & -0.6 & 73.5 \\
    \cline{2-9}\rule{0pt}{2.4ex}
    &\multirow{2}{*}{TF}&$\SELO$&-7.0    & 626.4 & -6.2  & 189.9 & -5.9 & 71.2 \\
    & &$\SELT$&-11.1 & 603.5 & -10.1 & 192.5 & -9.7 & 80.2 \\
    \cline{2-9}\rule{0pt}{2.4ex}
    &\multirow{2}{*}{FT}&$\SELO$&-0.8  & 703.5 & -0.5  & 209.9 & -0.6 & 76.2 \\
    & &$\SELT$&-1.3  & 699.9 & -0.8  & 210.7 & -0.6 & 74.9 \\
    \hline\rule{0pt}{2.4ex}
    \multirow{6}{*}{200}&\multirow{2}{*}{TT}&$\SELO$&-2.7  & 350.4 & -1.5  & 104.4 & -1.0   & 36.9 \\
    & &$\SELT$&-1.9  & 339.4 & -1.2  & 101.9 & -0.8 & 36.4 \\
    \cline{2-9}\rule{0pt}{2.4ex}
    &\multirow{2}{*}{TF}&$\SELO$&-5.8  & 326.8 & -4.6  & 99.5  & -4.1 & 37.6 \\
    & &$\SELT$&-7.8  & 320.9 & -6.8  & 105.9 & -6.3 & 45.6 \\
    \cline{2-9}\rule{0pt}{2.4ex}
    &\multirow{2}{*}{FT}&$\SELO$&-2.8  & 335.1 & -1.7  & 102.3 & -1.1 & 36.9 \\
    & &$\SELT$&-2.7  & 332.5 & -1.7  & 100.5 & -1.0   & 36.0   \\
    \hline\rule{0pt}{2.4ex}
    \multirow{6}{*}{400}&\multirow{2}{*}{TT}&$\SELO$&-2.0    & 158.1 & -1.2  & 46.9  & -0.7 & 16.7 \\
    & &$\SELT$&-2.2  & 157.2 & -1.2  & 47.0    & -0.7 & 16.7 \\
    \cline{2-9}\rule{0pt}{2.4ex}
    &\multirow{2}{*}{TF}&$\SELO$&-3.9  & 155.3 & -2.9  & 47.5  & -2.4 & 17.8 \\
    & &$\SELT$&-4.9  & 162.3 & -4.0    & 53.9  & -3.5 & 24.2 \\
    \cline{2-9}\rule{0pt}{2.4ex}
    &\multirow{2}{*}{FT}&$\SELO$&-2.1  & 158.8 & -1.2  & 47.8  & -0.7 & 17.1 \\
    & &$\SELT$&-2.2  & 157.1 & -1.2  & 47.3  & -0.7 & 16.9\\
    \hline
    \end{tabular}
    \label{Pt03}
\end{table}

\begin{table}[pt]
    \caption{$\%RB$ and $MSE(\times100)$ for Point Estimators when $t=0.5$}
    \centering
    \begin{tabular}{cclrrrrrr}
    \hline\rule{0pt}{2.1ex}
    \multirow{2}{*}{n}& \multirow{2}{*}{Scenario}&\multirow{2}{*}{Estimator}&\multicolumn{2}{c}{$\rho=0.3$}&\multicolumn{2}{c}{$\rho=0.5$}&\multicolumn{2}{c}{$\rho=0.7$}\\
    \cline{4-9}\rule{0pt}{2.1ex}
    & & &$\% RB$ & $MSE$ &$\% RB$ & $MSE$& $\% RB$ & $MSE$\\
    \hline\rule{0pt}{2.4ex}
        \multirow{6}{*}{100}&\multirow{2}{*}{TT}&$\SELO$&2.0    & 457.0   & 1.1  & 139.0   & 0.8  & 50.5  \\
    & &$\SELT$&2.5  & 453.8 & 1.4  & 137.8 & 0.7  & 50.5 \\
    \cline{2-9}\rule{0pt}{2.4ex}
    &\multirow{2}{*}{TF}&$\SELO$&0.1  & 448.5 & -1.0   & 136.6 & -1.5 & 50.6 \\
    & &$\SELT$&-2.0   & 450.2 & -3.1 & 142.0   & -3.7 & 56.0   \\
    \cline{2-9}\rule{0pt}{2.4ex}
    &\multirow{2}{*}{FT}&$\SELO$&1.9  & 455.2 & 1.1  & 138.4 & 0.6  & 50.4  \\
    & &$\SELT$&2.0    & 449.0   & 1.2  & 136.0   & 0.7  & 49.5 \\
    \hline\rule{0pt}{2.4ex}
    \multirow{6}{*}{200}&\multirow{2}{*}{TT}&$\SELO$&2.7  & 226.8 & 1.3  & 68.0    & 0.7  & 24.6  \\
    & &$\SELT$&2.7  & 225.1 & 1.4  & 67.8  & 0.7  & 24.5  \\
    \cline{2-9}\rule{0pt}{2.4ex}
    &\multirow{2}{*}{TF}&$\SELO$&1.2  & 222.7 & -0.1 & 67.3  & -0.8 & 24.6  \\
    & &$\SELT$&-0.6 & 222.1 & -1.8 & 69.4  & -2.4 & 27.7  \\
    \cline{2-9}\rule{0pt}{2.4ex}
    &\multirow{2}{*}{FT}&$\SELO$&2.7  & 226.2 & 1.4  & 68.1  & 0.7  & 24.6  \\
    & &$\SELT$&2.5  & 223.6 & 1.2  & 67.2  & 0.6  & 24.2 \\
    \hline\rule{0pt}{2.4ex}
    \multirow{6}{*}{400}&\multirow{2}{*}{TT}&$\SELO$&-0.3 & 121.1 & -0.2 & 36.6  & -0.1 & 13.3  \\
    & &$\SELT$&-0.4 & 121.0   & -0.2 & 36.6  & -0.1 & 13.3  \\
    \cline{2-9}\rule{0pt}{2.4ex}
    &\multirow{2}{*}{TF}&$\SELO$&-1.2 & 120.7 & -1.0   & 36.6  & -0.9 & 13.4  \\
    & &$\SELT$&-1.9 & 121.8 & -1.7 & 37.9  & -1.6 & 14.8 \\
    \cline{2-9}\rule{0pt}{2.4ex}
    &\multirow{2}{*}{FT}&$\SELO$&-0.5 & 121.0   & -0.3 & 36.7  & -0.1 & 13.3  \\
    & &$\SELT$&-0.5 & 119.6 & -0.2 & 36.3  & -0.1 & 13.2 \\
    \hline
    \end{tabular}
    \label{Pt05}
\end{table}

\begin{table}[pt]
    \caption{$\%RB$ and $MSE(\times100)$ for Point Estimators when $t=0.7$}
    \centering
    \begin{tabular}{cclrrrrrr}
    \hline\rule{0pt}{2.1ex}
    \multirow{2}{*}{n}& \multirow{2}{*}{Scenario}&\multirow{2}{*}{Estimator}&\multicolumn{2}{c}{$\rho=0.3$}&\multicolumn{2}{c}{$\rho=0.5$}&\multicolumn{2}{c}{$\rho=0.7$}\\
    \cline{4-9}\rule{0pt}{2.1ex}
    & & &$\% RB$ & $MSE$ &$\% RB$ & $MSE$& $\% RB$ & $MSE$\\
    \hline\rule{0pt}{2.4ex}
        \multirow{6}{*}{100}&\multirow{2}{*}{TT}&$\SELO$&-0.5 & 549.5 & -0.3 & 165.8 & -0.2 & 59.8 \\
    & &$\SELT$&-0.5 & 546.7 & -0.3 & 164.8 & -0.2 & 59.4 \\
    \cline{2-9}\rule{0pt}{2.4ex}
    &\multirow{2}{*}{TF}&$\SELO$&-1.6 & 545.8 & -1.4 & 166.0   & -1.3 & 60.7 \\
    & &$\SELT$&-2.9 & 543.1 & -2.9 & 165.9 & -2.8 & 61.7 \\
    \cline{2-9}\rule{0pt}{2.4ex}
    &\multirow{2}{*}{FT}&$\SELO$&-0.6 & 546.5 & -0.2 & 164.9 & -0.1 & 59.1 \\
    & &$\SELT$&-0.7 & 543.3 & -0.2 & 163.0   & -0.1 & 58.5 \\
    \hline\rule{0pt}{2.4ex}
    \multirow{6}{*}{200}&\multirow{2}{*}{TT}&$\SELO$&1.8  & 256.1 & 0.8  & 76.2  & 0.4  & 27.0   \\
    & &$\SELT$&1.7  & 256.3 & 0.8  & 76.3  & 0.4  & 27.0   \\
    \cline{2-9}\rule{0pt}{2.4ex}
    &\multirow{2}{*}{TF}&$\SELO$&1.1  & 256.8 & 0.2  & 76.6  & -0.3 & 27.3 \\
    & &$\SELT$&-0.1 & 257.1 & -1.0   & 77.5  & -1.4 & 28.4 \\
    \cline{2-9}\rule{0pt}{2.4ex}
    &\multirow{2}{*}{FT}&$\SELO$&1.4  & 253.3 & 0.7  & 75.3  & 0.3  & 26.7 \\
    & &$\SELT$&1.5  & 252.4 & 0.7  & 75.2  & 0.3  & 26.6 \\
    \hline\rule{0pt}{2.4ex}
    \multirow{6}{*}{400}&\multirow{2}{*}{TT}&$\SELO$&2.6  & 125.6 & 1.4  & 37.8  & 0.8  & 13.6 \\
    & &$\SELT$&2.6  & 125.7 & 1.4  & 37.8  & 0.8  & 13.6 \\
    \cline{2-9}\rule{0pt}{2.4ex}
    &\multirow{2}{*}{TF}&$\SELO$&2.3  & 125.8 & 1.2  & 37.9  & 0.6  & 13.7 \\
    & &$\SELT$&1.9  & 127.0   & 0.7  & 38.9  & 0.1  & 14.6 \\
    \cline{2-9}\rule{0pt}{2.4ex}
    &\multirow{2}{*}{FT}&$\SELO$&2.5  & 124.2 & 1.4  & 37.4  & 0.8  & 13.5 \\
    & &$\SELT$&2.4  & 123.7 & 1.3  & 37.2  & 0.7  & 13.4\\
    \hline
    \end{tabular}
    \label{Pt07}
\end{table}

\begin{table}[pt]\small
    \caption{$\%CP$ and $AL(\times100)$ for $95\%$ Confidence Intervals when $t=0.3$}
    \centering
    \begin{tabular}{cclrrrrrr}
    \hline\rule{0pt}{2.1ex}
    \multirow{2}{*}{n}& \multirow{2}{*}{Scenario}&\multirow{2}{*}{Estimator}&\multicolumn{2}{c}{$\rho=0.3$}&\multicolumn{2}{c}{$\rho=0.5$}&\multicolumn{2}{c}{$\rho=0.7$}\\
    \cline{4-9}\rule{0pt}{2.1ex}
    & & &$\% CP$ & $AL$ &$\% CP$ & $AL$& $\% CP$ & $AL$\\
    \hline\rule{0pt}{2.4ex}
    \multirow{12}{*}{100}&\multirow{4}{*}{TT}&$\mathcal{I}_{\mbox{\tiny SELR1}}$&92.7 & 1003.4 & 93.4 & 578.4 & 95.1 & 395.3 \\
    & &$\mathcal{I}_{\mbox{\tiny SELR1B}}$&97.4 & 1356.6 & 97.6 & 739.5 & 97.8 & 447.3 \\
    & &$\mathcal{I}_{\mbox{\tiny SELR2}}$&93.0   & 961.0    & 93.3 & 542.5 & 94.6 & 351.7 \\
    & &$\mathcal{I}_{\mbox{\tiny SELR2B}}$&97.0   & 1247.0   & 97.2 & 680.6 & 97.2 & 405.6 \\
    \cline{2-9}\rule{0pt}{2.1ex}
    &\multirow{4}{*}{TF}&$\mathcal{I}_{\mbox{\tiny SELR1}}$&92.5 & 944.7  & 93.4 & 543.3 & 94.1 & 360.5 \\
    & &$\mathcal{I}_{\mbox{\tiny SELR1B}}$&96.9 & 1172.9 & 96.6 & 648.6 & 96.5 & 398.9 \\
    & &$\mathcal{I}_{\mbox{\tiny SELR2}}$&93.0   & 916.5  & 92.7 & 512.6 & 91.6 & 325.4 \\
    & &$\mathcal{I}_{\mbox{\tiny SELR2B}}$&96.2 & 1074.1 & 96.3 & 616.0   & 95.8 & 407.5 \\
    \cline{2-9}\rule{0pt}{2.1ex}
    &\multirow{4}{*}{FT}&$\mathcal{I}_{\mbox{\tiny SELR1}}$&89.5 & 879.6  & 89.7 & 484.6 & 91.2 & 298.7 \\
    & &$\mathcal{I}_{\mbox{\tiny SELR1B}}$&97.6 & 1331.3 & 97.6 & 728.4 & 97.9 & 440.7 \\
    & &$\mathcal{I}_{\mbox{\tiny SELR2}}$&89.5 & 890.7  & 89.9 & 490.0   & 91.4 & 298.9 \\
    & &$\mathcal{I}_{\mbox{\tiny SELR2B}}$&97.0   & 1241.0   & 97.0   & 678.3 & 96.9 & 403.8 \\
    \hline\rule{0pt}{2.1ex}
     \multirow{12}{*}{200}&\multirow{4}{*}{TT}&$\mathcal{I}_{\mbox{\tiny SELR1}}$&92.0   & 694.9  & 92.3 & 402.4 & 94.7 & 267.4 \\
    & &$\mathcal{I}_{\mbox{\tiny SELR1B}}$&95.0   & 777.4  & 95.2 & 425.1 & 95.3 & 253.4 \\
    & &$\mathcal{I}_{\mbox{\tiny SELR2}}$&92.0   & 680.1  & 92.7 & 392.3 & 94.9 & 249.2 \\
    & &$\mathcal{I}_{\mbox{\tiny SELR2B}}$&94.6 & 761.2  & 94.7 & 416.9 & 95.0   & 248.9 \\
    \cline{2-9}\rule{0pt}{2.1ex}
    &\multirow{4}{*}{TF}&$\mathcal{I}_{\mbox{\tiny SELR1}}$&92.0   & 688.5  & 92.2 & 397.0   & 93.0   & 260.9 \\
    & &$\mathcal{I}_{\mbox{\tiny SELR1B}}$&94.4 & 724.9  & 93.6 & 399.0   & 93.4 & 243.1 \\
    & &$\mathcal{I}_{\mbox{\tiny SELR2}}$&91.9 & 662.4  & 91.7 & 373.8 & 89.6 & 236.0   \\
    & &$\mathcal{I}_{\mbox{\tiny SELR2B}}$&93.3 & 715.6  & 93.3 & 412.3 & 93.2 & 275.3 \\
    \cline{2-9}\rule{0pt}{2.1ex}
    &\multirow{4}{*}{FT}&$\mathcal{I}_{\mbox{\tiny SELR1}}$&91.0   & 642.3  & 90.7 & 353.5 & 91.2 & 213.7 \\
    & &$\mathcal{I}_{\mbox{\tiny SELR1B}}$&95.0   & 772.6  & 94.9 & 425.1 & 95.8 & 254.2 \\
    & &$\mathcal{I}_{\mbox{\tiny SELR2}}$&91.1 & 645.4  & 91.2 & 355.3 & 91.5 & 214.9 \\
    & &$\mathcal{I}_{\mbox{\tiny SELR2B}}$&94.6 & 758.3  & 94.7 & 416.0   & 94.6 & 248.1 \\
    \hline\rule{0pt}{2.1ex}
     \multirow{12}{*}{400}&\multirow{4}{*}{TT}&$\mathcal{I}_{\mbox{\tiny SELR1}}$&93.9 & 480.9  & 94.7 & 274.6 & 95.7 & 178.3 \\
    & &$\mathcal{I}_{\mbox{\tiny SELR1B}}$&95.7 & 505.1  & 96.0   & 276.0   & 96.3 & 164.7 \\
    & &$\mathcal{I}_{\mbox{\tiny SELR2}}$&94.3 & 483.0    & 95.2 & 274.2 & 96.1 & 172.8 \\
    & &$\mathcal{I}_{\mbox{\tiny SELR2B}}$&95.3 & 503.1  & 95.7 & 275.5 & 95.7 & 164.7 \\
    \cline{2-9}\rule{0pt}{2.1ex}
    &\multirow{4}{*}{TF}&$\mathcal{I}_{\mbox{\tiny SELR1}}$&94.6 & 479.8  & 95.1 & 273.5 & 95.6 & 176.4 \\
    & &$\mathcal{I}_{\mbox{\tiny SELR1B}}$&95.3 & 491.6  & 95.6 & 270.4 & 95.0   & 163.4 \\
    & &$\mathcal{I}_{\mbox{\tiny SELR2}}$&93.8 & 478.3  & 93.9 & 271.1 & 93.4 & 173.9 \\
    & &$\mathcal{I}_{\mbox{\tiny SELR2B}}$&95.4 & 505.6  & 95.4 & 294.9 & 94.9 & 198.6 \\
    \cline{2-9}\rule{0pt}{2.1ex}
    &\multirow{4}{*}{FT}&$\mathcal{I}_{\mbox{\tiny SELR1}}$&93.3 & 460.4  & 93.6 & 253.5 & 94.2 & 152.9 \\
    & &$\mathcal{I}_{\mbox{\tiny SELR1B}}$&95.8 & 508.7  & 96.0   & 279.6 & 95.8 & 167.0   \\
    & &$\mathcal{I}_{\mbox{\tiny SELR2}}$&93.1 & 458.6  & 93.5 & 252.5 & 94.1 & 152.3 \\
    & &$\mathcal{I}_{\mbox{\tiny SELR2B}}$&95.3 & 503.0    & 95.6 & 275.9 & 95.4 & 164.6\\
    \hline
    \label{CIt03}
    \end{tabular}
\end{table}

\begin{table}[pt]\small
    \caption{$\%CP$ and $AL(\times100)$ for $95\%$ Confidence Intervals when $t=0.5$}
    \centering
    \begin{tabular}{cclrrrrrr}
    \hline\rule{0pt}{2.1ex}
    \multirow{2}{*}{n}& \multirow{2}{*}{Scenario}&\multirow{2}{*}{Estimator}&\multicolumn{2}{c}{$\rho=0.3$}&\multicolumn{2}{c}{$\rho=0.5$}&\multicolumn{2}{c}{$\rho=0.7$}\\
    \cline{4-9}\rule{0pt}{2.1ex}
    & & &$\% CP$ & $AL$ &$\% CP$ & $AL$& $\% CP$ & $AL$\\
    \hline\rule{0pt}{2.4ex}
    \multirow{12}{*}{100}&\multirow{4}{*}{TT}&$\mathcal{I}_{\mbox{\tiny SELR1}}$&94.3 & 858.6 & 94.6 & 486.5 & 95.4 & 309.7 \\
    & &$\mathcal{I}_{\mbox{\tiny SELR1B}}$&96.9 & 946.6 & 96.5 & 520.1 & 96.1 & 308.0   \\
    & &$\mathcal{I}_{\mbox{\tiny SELR2}}$&94.4 & 831.9 & 94.8 & 469.0   & 95.6 & 296.5 \\
    & &$\mathcal{I}_{\mbox{\tiny SELR2B}}$&96.6 & 921.6 & 96.0   & 506.1 & 95.9 & 303.8 \\
    \cline{2-9}\rule{0pt}{2.1ex}
    &\multirow{4}{*}{TF}&$\mathcal{I}_{\mbox{\tiny SELR1}}$&94.6 & 852.2 & 95.3 & 492.5 & 95.8 & 330.2 \\
    & &$\mathcal{I}_{\mbox{\tiny SELR1B}}$&96.4 & 901.6 & 96.6 & 493.7 & 95.4 & 296.2 \\
    & &$\mathcal{I}_{\mbox{\tiny SELR2}}$&94.2 & 816.1 & 93.8 & 455.2 & 93.8 & 283.1 \\
    & &$\mathcal{I}_{\mbox{\tiny SELR2B}}$&95.6 & 882.3 & 95.7 & 498.6 & 96.0   & 318.7 \\
    \cline{2-9}\rule{0pt}{2.1ex}
    &\multirow{4}{*}{FT}&$\mathcal{I}_{\mbox{\tiny SELR1}}$&93.9 & 804.5 & 93.7 & 442.4 & 93.4 & 266.5 \\
    & &$\mathcal{I}_{\mbox{\tiny SELR1B}}$&97.0   & 932.8 & 96.7 & 512.5 & 96.2 & 307.6 \\
    & &$\mathcal{I}_{\mbox{\tiny SELR2}}$&93.7 & 804.2 & 93.7 & 441.5 & 93.2 & 266.1 \\
    & &$\mathcal{I}_{\mbox{\tiny SELR2B}}$&96.7 & 919.8 & 96.4 & 504.8 & 96.3 & 302.4 \\
    \hline\rule{0pt}{2.1ex}
     \multirow{12}{*}{200}&\multirow{4}{*}{TT}&$\mathcal{I}_{\mbox{\tiny SELR1}}$&95.2 & 589.7 & 95.7 & 331.8 & 96.0   & 210.7 \\
    & &$\mathcal{I}_{\mbox{\tiny SELR1B}}$&96.5 & 615.2 & 96.7 & 337.3 & 96.3 & 202.3 \\
    & &$\mathcal{I}_{\mbox{\tiny SELR2}}$&95.4 & 588.9 & 95.8 & 332.2 & 96.3 & 208.8 \\
    & &$\mathcal{I}_{\mbox{\tiny SELR2B}}$&96.4 & 611.5 & 96.5 & 335.7 & 96.2 & 201.4 \\
    \cline{2-9}\rule{0pt}{2.1ex}
    &\multirow{4}{*}{TF}&$\mathcal{I}_{\mbox{\tiny SELR1}}$&95.7 & 594.0   & 96.1 & 340.6 & 97.0   & 224.2 \\
    & &$\mathcal{I}_{\mbox{\tiny SELR1B}}$&96.0   & 605.2 & 95.9 & 331.9 & 95.6 & 198.9 \\
    & &$\mathcal{I}_{\mbox{\tiny SELR2}}$&94.7 & 583.5 & 94.2 & 325.5 & 94.2 & 202.8 \\
    & &$\mathcal{I}_{\mbox{\tiny SELR2B}}$&95.7 & 606.8 & 95.7 & 344.1 & 95.5 & 221.6 \\
    \cline{2-9}\rule{0pt}{2.1ex}
    &\multirow{4}{*}{FT}&$\mathcal{I}_{\mbox{\tiny SELR1}}$&94.9 & 573.0   & 94.9 & 315.1 & 95.1 & 189.8 \\
    & &$\mathcal{I}_{\mbox{\tiny SELR1B}}$&96.7 & 615.1 & 96.3 & 337.9 & 96.0   & 202.9 \\
    & &$\mathcal{I}_{\mbox{\tiny SELR2}}$&94.6 & 570.6 & 94.4 & 313.6 & 95.0   & 188.8 \\
    & &$\mathcal{I}_{\mbox{\tiny SELR2B}}$&96.7 & 610.9 & 96.7 & 335.4 & 96.3 & 201.0   \\
    \hline\rule{0pt}{2.1ex}
     \multirow{12}{*}{400}&\multirow{4}{*}{TT}&$\mathcal{I}_{\mbox{\tiny SELR1}}$&92.9 & 412.1 & 92.9 & 229.2 & 93.2 & 141.9 \\
    & &$\mathcal{I}_{\mbox{\tiny SELR1B}}$&93.7 & 421.9 & 93.2 & 231.2 & 93.0   & 138.8 \\
    & &$\mathcal{I}_{\mbox{\tiny SELR2}}$&93.0   & 412.5 & 92.9 & 229.4 & 93.2 & 142.1 \\
    & &$\mathcal{I}_{\mbox{\tiny SELR2B}}$&93.8 & 421.4 & 93.6 & 231.3 & 93.1 & 139.0   \\
    \cline{2-9}\rule{0pt}{2.1ex}
    &\multirow{4}{*}{TF}&$\mathcal{I}_{\mbox{\tiny SELR1}}$&93.1 & 416.8 & 93.9 & 237.1 & 95.1 & 153.9 \\
    & &$\mathcal{I}_{\mbox{\tiny SELR1B}}$&93.4 & 418.8 & 93.3 & 229.5 & 93.1 & 137.7 \\
    & &$\mathcal{I}_{\mbox{\tiny SELR2}}$&92.9 & 413.0   & 92.7 & 230.3 & 92.3 & 143.5 \\
    & &$\mathcal{I}_{\mbox{\tiny SELR2B}}$&93.8 & 423.9 & 93.7 & 239.6 & 93.7 & 154.0   \\
    \cline{2-9}\rule{0pt}{2.1ex}
    &\multirow{4}{*}{FT}&$\mathcal{I}_{\mbox{\tiny SELR1}}$&92.6 & 405.7 & 92.4 & 223.1 & 92.2 & 134.2 \\
    & &$\mathcal{I}_{\mbox{\tiny SELR1B}}$&93.9 & 422.0   & 93.5 & 231.5 & 93.2 & 139.0   \\
    & &$\mathcal{I}_{\mbox{\tiny SELR2}}$&92.6 & 403.8 & 92.7 & 221.9 & 92.4 & 133.5 \\
    & &$\mathcal{I}_{\mbox{\tiny SELR2B}}$&93.8 & 421.4 & 93.8 & 231.1 & 93.4 & 138.8\\
    \hline
    \label{CIt05}
    \end{tabular}
\end{table}

\begin{table}[pt]\small
    \caption{$\%CP$ and $AL(\times100)$ for $95\%$ Confidence Intervals when $t=0.7$}
    \centering
    \begin{tabular}{cclrrrrrr}
    \hline\rule{0pt}{2.1ex}
    \multirow{2}{*}{n}& \multirow{2}{*}{Scenario}&\multirow{2}{*}{Estimator}&\multicolumn{2}{c}{$\rho=0.3$}&\multicolumn{2}{c}{$\rho=0.5$}&\multicolumn{2}{c}{$\rho=0.7$}\\
    \cline{4-9}\rule{0pt}{2.1ex}
    & & &$\% CP$ & $AL$ &$\% CP$ & $AL$& $\% CP$ & $AL$\\
    \hline\rule{0pt}{2.4ex}
    \multirow{12}{*}{100}&\multirow{4}{*}{TT}&$\mathcal{I}_{\mbox{\tiny SELR1}}$&93.7 & 884.1  & 94.1 & 496.5 & 94.9 & 311.7 \\
    & &$\mathcal{I}_{\mbox{\tiny SELR1B}}$&96.5 & 1019.3 & 96.5 & 558.8 & 97.1 & 334.2 \\
    & &$\mathcal{I}_{\mbox{\tiny SELR2}}$&93.1 & 855.7  & 93.7 & 477.5 & 95.1 & 296.8 \\
    & &$\mathcal{I}_{\mbox{\tiny SELR2B}}$&95.8 & 997.1  & 95.9 & 545.1 & 96.3 & 326.7 \\
    \cline{2-9}\rule{0pt}{2.1ex}
    &\multirow{4}{*}{TF}&$\mathcal{I}_{\mbox{\tiny SELR1}}$&94.1 & 906.0    & 95.2 & 533.6 & 96.8 & 370.1 \\
    & &$\mathcal{I}_{\mbox{\tiny SELR1B}}$&96.3 & 1001.2 & 96.1 & 548.1 & 96.4 & 327.5 \\
    & &$\mathcal{I}_{\mbox{\tiny SELR2}}$&92.6 & 851.3  & 93.5 & 472.7 & 93.3 & 292.1 \\
    & &$\mathcal{I}_{\mbox{\tiny SELR2B}}$&95.7 & 967.5  & 95.6 & 537.0   & 96.3 & 334.5 \\
    \cline{2-9}\rule{0pt}{2.1ex}
    &\multirow{4}{*}{FT}&$\mathcal{I}_{\mbox{\tiny SELR1}}$&93.5 & 848.1  & 93.6 & 467.8 & 94.1 & 286.8 \\
    & &$\mathcal{I}_{\mbox{\tiny SELR1B}}$&96.5 & 1017.5 & 96.3 & 558.0   & 96.8 & 334.0   \\
    & &$\mathcal{I}_{\mbox{\tiny SELR2}}$&93.1 & 836.1  & 93.6 & 463.2 & 94.6 & 284.9 \\
    & &$\mathcal{I}_{\mbox{\tiny SELR2B}}$&95.8 & 992.3  & 96.1 & 543.3 & 96.4 & 325.7 \\
    \hline\rule{0pt}{2.1ex}
     \multirow{12}{*}{200}&\multirow{4}{*}{TT}&$\mathcal{I}_{\mbox{\tiny SELR1}}$&94.1 & 621.0    & 94.6 & 342.9 & 94.9 & 209.8 \\
    & &$\mathcal{I}_{\mbox{\tiny SELR1B}}$&95.3 & 646.3  & 95.7 & 354.4 & 95.4 & 212.4 \\
    & &$\mathcal{I}_{\mbox{\tiny SELR2}}$&93.9 & 613.0    & 94.7 & 339.7 & 95.3 & 208.2 \\
    & &$\mathcal{I}_{\mbox{\tiny SELR2B}}$&94.9 & 643.7  & 95.1 & 353.0   & 95.5 & 211.7 \\
    \cline{2-9}\rule{0pt}{2.1ex}
    &\multirow{4}{*}{TF}&$\mathcal{I}_{\mbox{\tiny SELR1}}$&94.4 & 633.9  & 96.0   & 368.1 & 97.7 & 248.8 \\
    & &$\mathcal{I}_{\mbox{\tiny SELR1B}}$&95.2 & 642.7  & 95.5 & 351.4 & 94.9 & 209.2 \\
    & &$\mathcal{I}_{\mbox{\tiny SELR2}}$&94.5 & 612.0    & 94.9 & 338.1 & 94.9 & 206.4 \\
    & &$\mathcal{I}_{\mbox{\tiny SELR2B}}$&95.3 & 640.3  & 95.8 & 354.9 & 95.8 & 218.5 \\
    \cline{2-9}\rule{0pt}{2.1ex}
    &\multirow{4}{*}{FT}&$\mathcal{I}_{\mbox{\tiny SELR1}}$&94.0   & 609.1  & 94.2 & 334.6 & 95.3 & 203.8 \\
    & &$\mathcal{I}_{\mbox{\tiny SELR1B}}$&95.6 & 649.6  & 96.2 & 356.3 & 95.8 & 213.5 \\
    & &$\mathcal{I}_{\mbox{\tiny SELR2}}$&93.9 & 600.9  & 94.5 & 331.7 & 95.0   & 202.2 \\
    & &$\mathcal{I}_{\mbox{\tiny SELR2B}}$&94.9 & 644.0    & 95.5 & 353.0   & 95.8 & 211.8 \\
    \hline\rule{0pt}{2.1ex}
     \multirow{12}{*}{400}&\multirow{4}{*}{TT}&$\mathcal{I}_{\mbox{\tiny SELR1}}$&94.7 & 437.7  & 94.8 & 242.4 & 94.6 & 148.4 \\
    & &$\mathcal{I}_{\mbox{\tiny SELR1B}}$&94.7 & 443.5  & 94.9 & 243.2 & 95.2 & 145.7 \\
    & &$\mathcal{I}_{\mbox{\tiny SELR2}}$&94.7 & 436.4  & 94.6 & 240.6 & 94.5 & 146.6 \\
    & &$\mathcal{I}_{\mbox{\tiny SELR2B}}$&94.9 & 443.4  & 94.8 & 243.1 & 95.1 & 145.8 \\
    \cline{2-9}\rule{0pt}{2.1ex}
    &\multirow{4}{*}{TF}&$\mathcal{I}_{\mbox{\tiny SELR1}}$&95.2 & 448.6  & 95.8 & 260.8 & 97.6 & 176.2 \\
    & &$\mathcal{I}_{\mbox{\tiny SELR1B}}$&94.8 & 442.4  & 94.9 & 242.1 & 94.5 & 144.5 \\
    & &$\mathcal{I}_{\mbox{\tiny SELR2}}$&94.5 & 435.5  & 94.7 & 240.9 & 94.3 & 147.2 \\
    & &$\mathcal{I}_{\mbox{\tiny SELR2B}}$&94.9 & 444.6  & 95.2 & 247.1 & 95.1 & 152.8 \\
    \cline{2-9}\rule{0pt}{2.1ex}
    &\multirow{4}{*}{FT}&$\mathcal{I}_{\mbox{\tiny SELR1}}$&94.2 & 429.9  & 94.6 & 237.0   & 94.7 & 143.7 \\
    & &$\mathcal{I}_{\mbox{\tiny SELR1B}}$&94.9 & 445.1  & 95.0   & 243.7 & 95.0   & 146.0   \\
    & &$\mathcal{I}_{\mbox{\tiny SELR2}}$&94.2 & 426.1  & 94.6 & 234.8 & 94.9 & 142.4 \\
    & &$\mathcal{I}_{\mbox{\tiny SELR2B}}$&95.0   & 443.6  & 95.2 & 242.9 & 95.5 & 145.7\\
    \hline
    \label{CIt07}
    \end{tabular}
\end{table}

\bigskip

\end{document}